\documentclass[a4paper,fleqn,usenatbib,useAMS]{mnras}

\usepackage{natbib}
\usepackage{url}
\usepackage{rotating}
\usepackage{subcaption}
\usepackage{amsmath}
\usepackage{amsfonts}
\usepackage{amssymb}
\usepackage{graphicx}
\usepackage{float}
\usepackage{aas_macros}
\usepackage{color}
\usepackage{ mathrsfs }

\def\gsim{\;\rlap{\lower 2.5pt
 \hbox{$\sim$}}\raise 1.5pt\hbox{$>$}\;}
\def\lsim{\;\rlap{\lower 2.5pt
   \hbox{$\sim$}}\raise 1.5pt\hbox{$<$}\;}

\usepackage{hyperref}
\hypersetup{
    pdfnewwindow=true,      
   colorlinks=true,       
    linkcolor=blue,          
    citecolor=blue,        
    filecolor=blue,      
    urlcolor=blue           
}

\begin{document}

\title[Relativistic Boost of SMBHBs]{Testing the relativistic Doppler boost hypothesis for supermassive black hole binary candidates}

\author[M. Charisi et al.]{Maria Charisi$^{1,2}$\thanks{mcharisi@caltech.edu}, Zolt\'an Haiman$^{2}$, David Schiminovich$^{2}$,  Daniel J.~D'Orazio$^{3}$\\
$^1$TAPIR, California Institute of Technology, Pasadena, CA 91125, USA\\
$^2$Department of Astronomy, Columbia University, New York, NY 10027, USA \\
$^3$Astronomy Department, Harvard University, Cambridge, MA 02138, USA}

\maketitle

\begin{abstract}
Supermassive black hole binaries (SMBHBs) should be common in galactic
nuclei as a result of frequent galaxy mergers. Recently, a large
sample of sub-parsec SMBHB candidates was identified as bright
periodically variable quasars in optical surveys. If the observed
periodicity corresponds to the redshifted binary orbital period, the
inferred orbital velocities are relativistic ($v/c\approx 0.1$). The
optical and UV luminosities are expected to arise from gas bound to
the individual BHs, and would be modulated by the relativistic Doppler
effect. The optical and UV light curves should vary in tandem with
relative amplitudes which depend on the respective spectral slopes.
We constructed a control sample of 42 quasars with aperiodic
variability, to test whether this Doppler colour signature can be
distinguished from intrinsic chromatic variability. We found that the
Doppler signature can arise by chance in $\sim$20\% ($\sim$37\%) of
quasars in the nUV (fUV) band. These probabilities reflect the limited
quality of the control sample and represent upper limits on how
frequently quasars mimic the Doppler brightness+colour variations.  We
performed separate tests on the periodic quasar candidates, and found
that for the majority, the Doppler boost hypothesis requires an
unusually steep UV spectrum or an unexpectedly large BH mass and
orbital velocity. We conclude that at most $\sim$1/3rd of these
periodic candidates can harbor Doppler-modulated SMBHBs.
\end{abstract}


\section{Introduction}
\label{section:Introduction}

It is well established that most, if not all massive galaxies harbor
supermassive black holes (SMBHs) in their nuclei. The mass of the
central BH is correlated with several properties of the host galaxy
(e.g., velocity dispersion, bulge luminosity, etc.), which suggests
that the SMBH and the host galaxy may co-evolve
\citep{Kormendy2013}. In the hierarchical structure formation model,
galaxies and quasars are built-up from smaller progenitors through
frequent mergers \citep{Haehnelt2002}. These mergers should result in
a supermassive black hole binary (SMBHB) at the center of the newly
formed galaxy, surrounded by significant amounts of gas
\citep{BarnesHernquist1992}.

In the post-merger galaxy, the orbit of the binary shrinks initially
due to dynamical friction and subsequently by scattering nearby stars
\citep{Begelman1980}.  At small (sub-pc) separations, three-body
interactions become less efficient, and the binaries may stall for a
significant fraction of the Hubble time (see,
e.g. \citealt{Colpi2014}).  In this regime, the ambient gas, which is
expected to settle into a circumbinary disc \citep{Barnes2002}, may
dominate the binary's orbital decay \citep{Tang+2017}, while, at the
same time, it can accrete onto the BHs providing bright
electromagnetic emission (e.g. \citealt{2009ApJ...700.1952H}).  The
SMBHB is eventually driven to merger by the emission of low-frequency
gravitational waves (GWs), which could be detectable in the future by
pulsar timing arrays (PTAs; \citealt{iPTA}) and by the space-based
interferometer LISA~\citep{eLISA2017}.

The orbital decay of the binary due to its interaction with the
circumbinary disc is expected to be slow
\citep{2009ApJ...700.1952H,2012MNRAS.427.2680K,2012MNRAS.427.2660K,2013ApJ...774..144R,2016arXiv160205206R,Kelley+2017}
and SMBHBs should spend significant time ($\gsim 10^6$ years) at
sub-pc separations and thus should be fairly common. Despite their
expected abundance, SMBHBs have only been detected at large
separations, from several kpc
\citep{2003ApJ...582L..15K,2008MNRAS.386..105B,2010ApJ...710.1578G,2011ApJ...735L..42K,2011Natur.477..431F,2011ApJ...737L..19C,2015ApJ...799...72F}
down to a few pc \citep{2006ApJ...646...49R}. At sub-pc separations,
it is much more difficult to spatially resolve the orbit of the binary (see, however, \citealt{Dorazio_2017_VLBI})
 and one has to rely on the effects of the
binary on its environment. Several candidates have been identified
from large velocity offsets in quasar spectra, helical morphology of
radio jets, etc (see, e.g., recent reviews by \citealt{Dotti+2012} and
\citealt{KomossaZensus2016}).

SMBHBs can also be naturally identified as quasars with periodic
variability. First, in hierarchical structure formation models,
quasars are thought to be activated by galaxy mergers
\citep{KauffmanHaehnelt2000}.
Recent observations of interacting galaxies have provided further
evidence for excess AGN activity~\citep{Goulding+2017}.  Many quasars
may thus host SMBHBs. Second, numerous hydrodynamical simulations of
SMBHBs with circumbinary gas discs have found that such binaries would
produce bright quasar-like luminosities, but with the accretion rate
modulated periodically at the orbital period of the binary
\citep{AL96,2007PASJ...59..427H,2008ApJ...672...83M,2012ApJ...755...51N,2012A&A...545A.127R,2013MNRAS.436.2997D,Farris2014,2014PhRvD..89f4060G}. In
a binary system, periodic variability can also arise from a precessing
jet, as the viewing angle of the jet varies periodically~(see,
e.g. \citealt{Kun2014,Kun2015}).

Recently, a large number of quasars with significant periodicity has
emerged, mainly from systematic searches in the photometric databases
of large time-domain optical surveys. \citet[][hereafter
  G15]{Graham2015} identified 111 candidates in a sample of
$\sim$245,000 quasars from the Catalina Real-Time Transient Survey
(CRTS). \citet[][hereafter C16]{Charisi2016} analyzed a sample of
$\sim$35,000 quasars from the Palomar Transient Factory (PTF) and
identified 33 additional candidates, with typically shorter periods
and dimmer magnitudes. Another recent candidate, quasar SDSS
J0159+0105, was identified to have two periodic components in its
variability with a frequency ratio
1:2 from a smaller sample of
$\sim$ 350 low-redshift quasars in CRTS
\citep[]{Zheng2016}.\footnote{This quasar was also included in the
  sample analyzed in G15, but was not identified as periodic. The
  shortest periodic component of $\sim$740\,d was identified with both
  the wavelet and the autocorrelation function analysis in G15, but
  did not have enough power to be classified as periodic, potentially
  due to the second periodic component (M.~Graham; private
  communication).}

Identifying periodic variability in quasars is extremely challenging,
because quasars show stochastic variability. The overall stochastic
variability is successfully described by a damped random walk (DRW)
model
\citep{2009ApJ...698..895K,2010ApJ...708..927K,2010ApJ...721.1014M}
although some recent studies have suggested that a more advanced
description of quasar variability may be required
\citep{2011ApJ...743L..12M,2013ApJ...765..106Z,2014MNRAS.439..703G,2016A&A...585A.129S}.\footnote{The
  periodicity search algorithms and statistical analyses in these
  papers differ significantly. However, in all cases, the underlying
  assumption for the quasar variability is the DRW model.}
Additionally, the optical light curves are typically sparse, with
seasonal gaps and relatively short baselines compared to the periods
(the identified candidates were typically observed only for a limited
number of 2-3 cycles). As pointed out by \citet{Vaughan2016}, our
incomplete knowledge of quasar variability, in combination with the
limited quality of the optical light curves can lead to false
detections of periodicity. This was demonstrated in the case of the
quasar PSO J3334.2028+01.4075 \citep{Liu2015}, whose follow-up
monitoring failed to show persistent periodicity \citep{Liu2016}.
In addition, \citet{Sesana2017} calculated the GW background for the
population of SMBHBs implied by the identified periodic binary
candidates, and found that it is in tension with the current upper
limit derived from PTAs. The tension can be alleviated only if the
typical masses and/or the mass-ratios in this sample are surprisingly
lower than expected; this suggests that many of the candidates cannot
be genuine SMBHBs.

It is therefore crucial to explore additional signatures that could
support the binary hypothesis for any periodic candidate. Several such
studies followed the discovery of the first quasar with periodic
variability (PG 1302-102; \citealt{Graham2015Nature}), including the
search for multiple periodic components in the optical variability, as
expected from hydrodynamical simulations
\citep{Charisi2015,Dorazio2015}, the analysis of the helical structure
in the radio jet of PG 1302-102 \citep{Kun2015,Mohan2016}, and the
detection of periodic variability in the infrared
\citep{Jun2015,Dorazio2017}.

Another proposed signature of SMBHBs is to detect evidence for the
relativistic Doppler boost. Assuming that the observed periodicity
corresponds to the redshifted orbital period of the binary, we can
infer that most of the candidates are at sub-pc separations, orbiting
with mildly relativistic velocities (a few percent of the speed of
light). If the optical emission arises in gas bound to the individual BHs,
 e.g., in mini-discs seen in hydrodynamical simulations
 \citep{Farris2014,Ryan2017}, the luminosity of the brighter
 mini-disc, typically the circum-secondary disc, will be inevitably
 Doppler boosted. For near-equal-mass binaries, this
 Doppler-induced variability is expected to be sub-dominant to
 hydrodynamically-introduced fluctuations~\citep{Bowen_2017b,Bowen_2017,Tang+2018},
 but for unequal-mass binaries, it could dominate the variability~\citep{Dorazio+2016}.

The observed frequency of the emitted photons will change due to the
relativistic motion, whereas the number of photons, which is
proportional to $F_{\nu}/\nu^3$, with $F_{\nu}$ the flux at a specific
frequency $\nu$, is Lorentz invariant.  The photons will be Doppler
boosted by a factor
\begin{equation}
\label{eq:DopplerFactor}
 \mathcal{D} = \frac{1}{\gamma\left[ 1 - \frac{v_{\parallel}}{c}
     \right]}; \qquad \gamma \equiv \frac{1}{\sqrt{1 -
     \left(\frac{v}{c}\right)^2 }},
\end{equation}
where $v$ is the orbital velocity and $v_{\parallel}$ is its
line-of-sight component. Assuming that the emitted radiation has a
power-law spectrum $F_{\nu}\propto \nu^{\alpha_{\nu}}$, the observed
flux will be
\begin{equation}
F_{\nu}^{\rm obs}=D^{3-\alpha_{\nu}}  F_{\nu}^{\rm em}.
\end{equation}
For a binary on a circular orbit, it can be shown that the variability
due to Doppler boost to first order in $v/c$ is
\begin{equation}
\label{eq:boostAmplitude}
\frac{\Delta F_{\nu}}{F_{\nu}}=(3-\alpha_{\nu})\frac{v}{c}\cos\phi\sin i,
\end{equation}
where $v$ is the orbital velocity of the more luminous BH (typically
the less massive secondary BH, with the primary assumed to contribute
negligible flux), $i$ is the inclination of the orbit to the
line-of-sight and $0\leq\phi\leq2\pi$ is
the phase of the orbit.

Therefore, even if the optical luminosity in the mini-discs is
constant, the unresolved binary will appear blue-shifted (and
brighter, for typical spectral slopes, i.e. $\alpha_{\nu}<3$),
when the more luminous BH is
moving towards the observer, and vice-versa.  The relativistic Doppler
boost may naturally explain the observed light curves, which show
smooth quasi-sinusoidal periodicity. We note that the periodic mass
accretion rates found in hydrodynamical simulations listed in the
Introduction are more bursty than sinusoidal, and are expected to
produce more bursty light curves.

The relativistic Doppler boost provides a uniquely robust prediction,
which can be tested with multi-wavelength data. For instance, if the
UV emission also arises in the mini-discs, then the UV variability
should also follow eq. (\ref{eq:boostAmplitude}). This means that the
UV light curve should track the optical, but with a different
variability amplitude, which depends on the spectral indices in the
respective bands. If the spectrum follows power-laws in both bands
with spectral indices $\alpha_{\rm opt}$ and $\alpha_{\rm UV}$, the
relative amplitude follows from eq. (\ref{eq:boostAmplitude}),
\begin{equation}
\label{eq:DooplerBoost}
\frac{A_{\rm UV}}{A_{\rm opt}}=\frac{3-\alpha_{\rm UV}}{3-\alpha_{\rm opt}},
\end{equation}
where, $A_{\rm opt}$ and $A_{\rm UV}$ are the amplitudes of the
optical and UV variability.\footnote{The X-ray luminosity also arises
  very close to SMBHs and in the binary scenario, it may be Doppler
  boosted. In this paper, we will focus only on the optical/UV
  variability, because the available X-ray data are insufficient to
  extend our study to this band.}

\citet{Dorazio2015Nature} proposed this model to explain the
variability of the periodic candidate PG 1302-102. In particular, they
found that if PG 1302-102 hosts an unequal mass binary ($q\lsim
0.05$), orbiting not too far from edge-on ($\lsim 30^\circ$), the
Doppler boost should dominate the variability. The model can
successfully fit the observed optical light curve. Additionally, with
archival data from the Galaxy Evolution Explorer (GALEX) and the
Hubble Space Telescope (HST), they showed that the variability
amplitudes in the near-UV (nUV) and far-UV (fUV) bands are
$A_{\rm nUV}/A_{\rm opt}\sim2.13$ and $A_{\rm fUV}/A_{\rm opt}\sim2.63$,
respectively, consistent with the prediction from
eq.~(\ref{eq:DooplerBoost}).

On the other hand, quasars are known to be variable across the
electromagnetic spectrum. The optical and UV luminosities typically
change almost simultaneously, with minimal interband time-lags
\citep{Cutri1985,Edelson1996,Giveon1999,Sakata2011}. Additionally,
several studies have indicated that quasars become
``bluer-when-brighter" (i.e. the continuum spectral slope becomes
steeper in brighter phases); this implies that the variability is
wavelength-dependent, with higher variability amplitudes at shorter
wavelengths \citep{Kinney1991, Paltani1994, VandenBerk2004,
  Wilhite2005, Ruan2014, Hung2016}. For instance, the variability
amplitudes in the UV are significantly larger than at optical
wavelengths \citep{Welsh2011,Zhu2016}.

The above trends suggest that, in general, quasars can show optical/UV
variability that may mimic the multi-wavelength variability predicted
by the Dopler model (eq.~\ref{eq:DooplerBoost}). In this paper, we
construct a control sample of aperiodic quasars, and determine how
often the intrinsic multi-wavelength variability of quasars in this
sample produces by random chance amplitudes consistent with the
predictions of the Doppler boost model.  Such a test is especially
important given that the currently available UV data are typically
relatively sparse.  More specifically, we analyze optical and UV light
curves to determine the ratio of the observed variability amplitudes
in the two bands (i.e. the left side of
eq.~\ref{eq:DooplerBoost}). Next, we fit the available optical and UV
spectra with power-laws, and from the inferred spectral indices, we
compute the expected UV-to-optical amplitude ratio (i.e. the right
side of eq.~\ref{eq:DooplerBoost}). We check whether the above two
quantities are consistent within their errors.  Additionally, as a
separate test, for a subsample of periodic quasars we estimate the UV
spectral index required in order for the putative binaries to be
Doppler boosted, and compare this to the range of observed UV spectral
slopes in the control sample.

This paper is organized as follows: In \S~\ref{section:SampleMethods},
we present the construction of the control sample, as well as the
sample of periodic candidates, and describe our data analysis. In
\S~\ref{Section:Results} we show the results of our analysis, followed
by a discussion in \S~\ref{sec:discussion}. We end with a short
summary of our findings and the implications of our results in
\S~\ref{sec:summary}.

\section{Sample and Methods}
\label{section:SampleMethods}

\subsection{Sample}
\label{section:Sample}

In order to statistically assess the significance of an apparent
Doppler signature,\footnote{For this test, the Doppler signature
  refers to the multi-wavelength variability amplitudes described in
  eq. (\ref{eq:DooplerBoost}). However, we emphasize that the Doppler
  model does not refer only to the relative amplitudes, but it can
  explain the overall variability, e.g., for a circular binary orbit,
  it can explain the sinusoidal light curve.}
our null hypothesis is that this signature can arise from intrinsic
wavelength-dependent variability of quasars, unrelated to binary black
holes. In other words, we test how often the relative amplitudes of
optical and UV variability are by chance consistent with the
prediction from eq. (\ref{eq:DooplerBoost}).  For this purpose, we
assembled a control sample of quasars and active galactic nuclei
(AGNs) that do not exhibit periodic variability, but whose properties
(luminosities and redshifts) resemble those of the periodic
candidates.
In order to perform this Doppler null test, the following data are necessary
for each source in the control sample: (1) an optical light curve, (2)
a UV light curve, temporally coincident with the optical,\footnote{We
  only considered UV light curves consisting of at least two distinct
  epochs, separated by 100\,days within the optical baseline (see
  below).} (3) an optical spectrum, and (4) a UV spectrum. Since the
availability of UV spectroscopy is limited, we maximized the sample
size, starting from a sample of sources with available UV
spectra.\footnote{Light curves in optical and UV bands, as well as
  optical spectra are available for very large samples of quasars from
  CRTS, GALEX and SDSS, respectively.}

More specifically, we made use of two HST spectroscopic catalogs,
which provide calibrated co-added UV spectra: (A) the Atlas of
Recalibrated HST Faint Object Spectrograph (FOS) Spectra of AGN and
Quasars \citep{Evans2004} and (B) the Cosmic Origin Spectrograph (COS)
quasar catalog of fUV spectra, provided by the Hubble Spectroscopic
Legacy Archive (HSLA; \citealt{HLSA}).\footnote{To the best of our
  knowledge, the above are the only available catalogs with high-level
  quasar and AGN spectra. Calibrated nUV spectra from COS, and from
  the Space Telescope Imaging Spectrograph (STIS) will be released in
  the future in the HSLA (\url{https://hla.stsci.edu}).} The FOS and
COS catalogs include spectra from 204 and 564 unique sources,
respectively, among which 56 are common in the two catalogs.  
 
From the COS catalog, we eliminated five sources that we could not
cross-correlate in the astronomical database
Simbad\footnote{\url{http://simbad.u-strasbg.fr/simbad/}} and two
additional sources that were classified as X-ray binaries. We also
excluded from both catalogs sources classified as BL Lac Objects,
since they are known to have distinct variability properties from
quasars; e.g. their variability may not arise in the accretion disc,
but in a relativistic jet \citep{Edelson1992}. These sources were also
excluded from the searches for periodicity in both G15 and C16. We
excluded five additional sources from the COS catalog, and one source
(PG 1302-102) which was included in both catalogs, because they were
identified as periodic.

For the remaining sources, we extracted optical spectra from the Sloan
Digital Sky Survey (SDSS),\footnote{\url{https://dr13.sdss.org}}
optical light curves from
CRTS\footnote{\url{http://nesssi.cacr.caltech.edu/DataRelease/}} and
UV light curves from GALEX.\footnote{\url{http://galex.stsci.edu}}
Since, as mentioned above, light curves and spectra in both bands are
necessary for the Doppler test we kept in the sample only the sources
that had all the necessary information.  We note that GALEX obtained
simultaneously photometric measurements with two UV filters (covering
the nUV and fUV bands), but the fUV light curves are typically more
sparse, because quasars are fainter in fUV. Moreover, the UV spectra
usually do not extend to both bands.  So if, for instance, the UV
spectrum covers only the fUV band, we checked only for the
availability of the fUV light curve (with the additional requirement
mentioned before, i.e. two distinct epochs and temporal coincidence
with the optical data).
We note that this step dramatically decreases the sample size; from
528 and 189 sources, in the COS and FOS catalogs (after removing
periodic sources and blazars), only 97 and 44 are left after this
cut. The main limitation is the availability of UV light curves and
even more so the coincidence of those light curves with the optical
baseline.

In order to obtain reliable estimates of the spectral slopes, we
imposed a redshift cut at $z\leq0.5$ for the sources examined in the
fUV band (sources both from the FOS and COS catalogs) and at
$z\leq1.3$ for the nUV sample (consisting only of sources from the FOS
catalog).  The redshift cut ensures that the UV continuum is not
significantly affected by intergalactic absorption \citep{Madau1996}.
In the fits, we did not consider wavelengths shorter than the Lyman
limit.
We also excluded nine sources with spectra of very poor quality or
with very strong absorption features, since it is difficult to
reliably estimate the spectral slopes from those.
 
Since we estimated the spectral slopes based only on the line-free
regions of the spectra (see below), we further required that the
wavelength range of those regions covers at least 25\% of the
respective GALEX band. As for the GALEX bands, we considered the
wavelength range at FWHM of the filter transmission curve extended by
10\% at both sides. This choice is dictated by our expectations for
the periodic sample; the observed variability amplitude is typically
$\sim$10\% and is consistent with the putative Doppler boost for the
inferred parameters of the binaries.

Given all the above necessary constraints, the test of the null
hypothesis was feasible for 42 sources (13 from the COS catalog, 30
from FOS, with one included in both catalogs). The test was feasible
in the fUV band for all the sources in the COS catalog and for four of
the FOS sources.\footnote{For these last four sources, the test was feasible both in the fUV and nUV bands.} 
In the nUV band, the test was performed with the 30
sources from the FOS catalog. In Table~\ref{Table:Sample_Properties},
 we summarize the number of sources included in each band.

\subsection{Optical and UV Spectra}

We assume that the continuum of quasar/AGN spectra can be approximated
by a single power-law $F_{\lambda}\sim\lambda^{\beta_{\lambda}}$,
where $\beta_{\lambda}=-\alpha_{\nu}-2$.  In order to estimate the
spectral index, it is essential to avoid the broad emission lines and
fit the power-law continuum to the line-free regions of the spectrum,
which are known to be significantly less variable than the continuum
\citep{Wilhite2005}. We also emphasize that in the case of SMBHBs, the
broad emission lines are likely produced in the circumbinary disc and
not in the mini-discs around the individual BHs \citep{Lu+2016}, where
the luminosity is expected to be Doppler boosted.

For the optical spectra, we automated the fit taking advantage of the
spectral line information provided by the SDSS pipeline. More
specifically, each line is fit with a Gaussian profile, and for lines
with a valid fit (as indicated by the pipeline), we removed the main
part of the line by interpolating between $-\sigma$ to $+\sigma$ from
the central wavelength. Subsequently, we smoothed the spectra (to
remove potential contribution of the parts of the broad lines that
were not removed by the interpolation) by calculating the moving
average over a wide window of 400 wavelength bins and fit a power-law
to the moving average.

For the UV spectra, on the other hand, the spectral line information
is not included in the catalogs. Therefore we identified the line free
regions manually, and fit the continuum based on the line-free
regions.  In both cases, we confirmed the validity of the fit by
visual inspection.

If more than one spectrum was available, we used the average slope
from all of the spectra. The variability in the spectral slope is
demonstrated in the Fig.~\ref{Fig:PeriodicSample_nUV}
and~\ref{Fig:PeriodicSample_fUV}, with those sources with large
horizontal error bars indicating large spectral-slope variability (see
also \ref{section:ConstantSpectralSlope} below).

\subsection{Photometric Fits}

We extracted the optical light curves from the public CRTS database
(Data Release 2). The survey combines data from three different
telescopes and thus the light curves may consist of multiple data
streams, which are not calibrated as a single light curve. In order to
avoid systematic effects, we selected the single light curve with the
highest number of observations. Additionally, the CRTS light curves
contain four data points per night (four observations per visit
separated by 10\,min). Since the short-term variability is not
significant for this study, we binned the observations taken within
the same night. Next, we extracted UV light curves from GALEX. These
light curves, especially in the fUV band, are typically sparse with
only a handful of epochs.

\begin{figure}
\includegraphics[height=6cm,width=8cm]{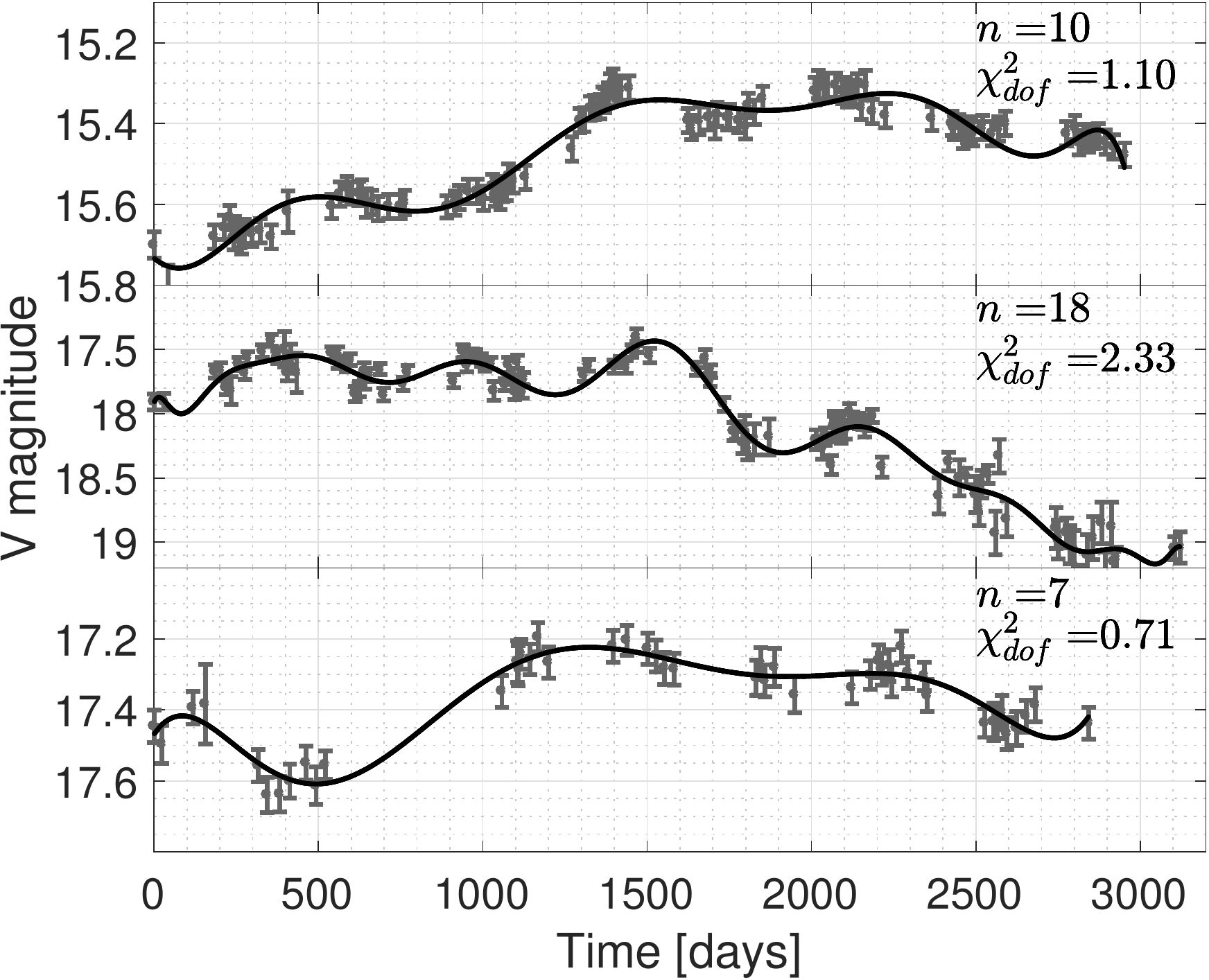}
\caption{Examples of optical light curves and the chosen
  polynomical fits with $\chi^2_{\rm dof}\sim1$ (top panel),
  $\chi^2_{\rm dof}>1$ (middle panel), and $\chi^2_{\rm dof}<1$
  (bottom panel). The degree of the polynomial $n$ is also shown in
  each panel.}
\label{Fig:PolyFit}
\end{figure}

The next step consisted of measuring the relative amplitudes in the
optical and UV bands (left side of eq. \ref{eq:DooplerBoost}). This
step is challenging for two main reasons: (1) quasars exhibit
stochastic variability, and (2) the optical and UV data were not taken
simultaneously. To address these issues, we approximated the optical
variability with an $n^{\rm th}$ degree polynomial. We chose the order
of the polynomial between 5 and 20 based on the reduced $\chi^2$ (we
chose the smallest polynomial order which gives $\chi^2_{\rm dof}\sim1$)
combined with visual inspection of each fitted light curve, to ensure
that the chosen fit reasonably represents the variability. For
instance, in some cases, the reduced $\chi^2$ is smaller than unity
($\chi^2_{\rm dof}<1$) for all the tested polynomials, which may suggest
that the provided error bars are overestimated, whereas, in other
cases, it is always higher than unity ($\chi^2_{\rm dof}>1$), because the
polynomial fit cannot entirely capture the short-term variability.  In
Fig.~\ref{Fig:PolyFit}, we show three optical light curves with the
chosen polynomial fit, with $\chi^2_{\rm dof}\sim1$ (top panel),
$\chi^2_{\rm dof}>1$ (middle panel), and $\chi^2_{\rm dof}<1$ (bottom
panel). The order of the chosen polynomial is also shown in each
panel.

We fit the UV light curve by rescaling the optical fit and shifting it
by a constant,
\begin{equation}
y_{\rm uv}^n=\theta_1\times y_{\rm opt}^n+\theta_2.
\end{equation}
Note that $y_{\rm uv}^i\equiv y_{\rm uv}(t^i)$, where $t^1, t^2, ... t^n$
represent the time series of the UV measurements, and $y_{\rm
  opt}(t^n)$ are the optical light-curves interpolated to these times
using the polynomial fits.
This can be re-written in matrix notation as
 \begin{equation}
Y_{\rm uv}=M\times \Theta,
\end{equation}
 where
 \begin{equation}
Y_{\rm uv}=\begin{bmatrix}
y_{\rm uv}(t^1)\\
y_{\rm uv}(t^2)\\
\vdots\\
y_{\rm uv}(t^n)
\end{bmatrix},
 \hspace{4mm}
  M=\begin{bmatrix}
  y_{\rm opt}(t^1)&1& \\
  y_{\rm opt}(t^2)&1&\\
  \vdots & \vdots & \\
  y_{\rm opt}(t^n)&1&
  \end{bmatrix},
  \hspace{4mm}
\Theta=\begin{bmatrix}
\theta_1\\
\theta_2
\end{bmatrix}.
 \end{equation}
 The maximum likelihood solution for this regression is
 \begin{equation}
\Theta=\left(M^T\times C^{-1}\times M\right)^{-1} \times\left(M^T \times C^{-1}\times Y_{\rm uv}\right)
 \end{equation}
 where $C$ is the covariance matrix of the UV measurement errors.

We emphasize that the polynomial fit provides a reasonable way for
interpolating the light curves, but it does not have predictive power
outside the observed time interval. Therefore, as mentioned above, UV
data points that do not overlap with the optical were excluded from
the fit.

\subsection{Periodic Sample}

We extracted UV light curves and optical spectra for the sample of
periodic quasars from G15 and C16. As mentioned above, with the
exception of PG 1302-102, which was analyzed in
\citet{Dorazio2015Nature}, only five of the periodic sources in G15
have UV spectra. Among those, four do not have multiple UV
observations from GALEX and one has multiple observations in nUV,
whereas the spectrum covers the fUV band. Therefore, directly testing
the Doppler model was not possible for any of the periodic sources.
However, for the sub-sample of sources that have UV light curves and
optical spectra (55 out of 111 sources in G15 and 13 out of 33 sources
in C16), we were still able to assess whether the Doppler boost model
is feasible.

More specifically, assuming that the optical and UV variability are
both dominated by the effects of relativistic Doppler boost, we
inferred the implied UV spectral index from
eq.~(\ref{eq:DooplerBoost}).  We were able to estimate the nUV
spectral index for 68 sources and for 27 of these sources we were also
able to calculate the fUV spectral index, see also Table~\ref{Table:Sample_Properties}. 
For this, we followed the
steps described above for fitting the optical spectra with a power-law
(in the $V$ band for CRTS and the $R$ band for PTF) and finding the
relative amplitude of UV and optical variability.\footnote{Since the
  optical light curves of the periodic sources are described by
  sinusoids, when fitting the UV light curves, we included UV data
  points outside of the optical baseline. This was not possible for
  the stochastic variability of quasars in the control sample.}
Subsequently, we compared our estimates including the 1-$\sigma$
uncertainties to the distribution of measured slopes from the control
sample to assess whether the inferred spectral slopes correspond to
realistic values seen in quasars with similar properties.

As we mentioned above, the Doppler signature is not limited only to
the multi-wavelength prediction from eq. (\ref{eq:DooplerBoost}). If
the emission of the periodic candidates is indeed due to relativistic
Doppler boost, the model should explain the entire sinusoidal
variability of the optical light curve as well.  For the candidates
that have at least one optical spectrum (17 from the PTF sample and 94
from the CRTS sample), we additionally checked whether the Doppler
model is feasible based only on the optical variability. To perform
this check, we computed the ratio of the maximum possible Doppler boost amplitude (for the
given parameters of the putative binaries) to the observed optical variability
amplitude.

In more detail, as is obvious from
eqs.~(\ref{eq:DopplerFactor})~and~(\ref{eq:boostAmplitude}), the
maximum Doppler boost occurs when the line-of-sight velocity is
maximum, which in turn occurs when the binary orbit is edge-on,
i.e. $\sin i=\sin(\pi/2)$. Additionally, the orbital velocity of the
secondary in a binary system depends on the mass ratio $q$,
\begin{equation}
v_2=\left(\frac{1}{1+q}\right) \left(2 \pi \frac{GM}{P}\right)^{1/3},
\end{equation}
and the maximum Doppler boost corresponds to this orbital velocity in
the limit of a very unequal mass binary (i.e. $q \to 0$). In this most
optimistic case, the Doppler factor is maximum (minimum), when the
secondary moving towards (away from) the observer,
\begin{equation}
 \mathcal{D}_{\max} = \frac{1}{\gamma\left[ 1 - \left(2 \pi\frac{GM}{c^3 P}\right)^{1/3} \right]},
\end{equation}
\begin{equation}
 \mathcal{D}_{\min} = \frac{1}{\gamma\left[ 1 + \left(2 \pi\frac{GM}{c^3 P}\right)^{1/3} \right]}.
\end{equation}
For each SMBHB candidate, the orbital period $P$ is known (assuming
that the observed period is the redshifted orbital period). The total
mass $M$ was either measured from the width of the broad emission
lines or estimated from the quasar's luminosity as in C16. The optical
spectral slope $\alpha_{\rm{\rm opt}}$, and the observed variability
half-amplitude
in magnitudes, $A_{\rm opt}$, were measured directly from the optical
spectra and the light curves, respectively.

Based on the above, the ratio of the
maximum possible range under the Doppler hypothesis to the observed variability range is given by
\begin{align}
\label{eq:FoM}
\frac{ \left. A_{\rm{Dop}} (M,P,\alpha_{\rm{\rm opt}})\right|_{\max}}{2 A_{\rm opt}} &= 
\frac{- 2.5 \log_{10} \left[F_{\rm opt}^{\max}/F_{\rm opt}^{\min} \right]}{2 A_{\rm opt}}\\
&=\frac{ - 2.5 \log_{10}\left( \mathcal{D}_{\max}/\mathcal{D}_{\min} \right)^{3-\alpha_{\rm{\rm opt}}} }{2 A_{\rm opt}}. \nonumber
\end{align}
If this ratio is below unity for a given SMBHB candidate, the
periodicity associated with that candidate cannot be generated solely
by relativistic Doppler boost.

\begin{table}
\caption{Number of sources with available data in the different bands in the control and periodic sample.}
\label{Table:Sample_Properties}
\begin{tabular}{c| c| c| c| c|}
&Optical& nUV & fUV &nUV+fUV\\
 \hline
 Control Sample&42&30&16&4\\
Periodic Sample&68&68&27&27\\
\end{tabular}
\end{table}

\section{Results}
\label{Section:Results}

\subsection{Null Hypothesis Test in Control Sample}
\label{section:Results_NullTest}

\begin{figure}
\centering
\includegraphics[height=6cm,width=9cm]{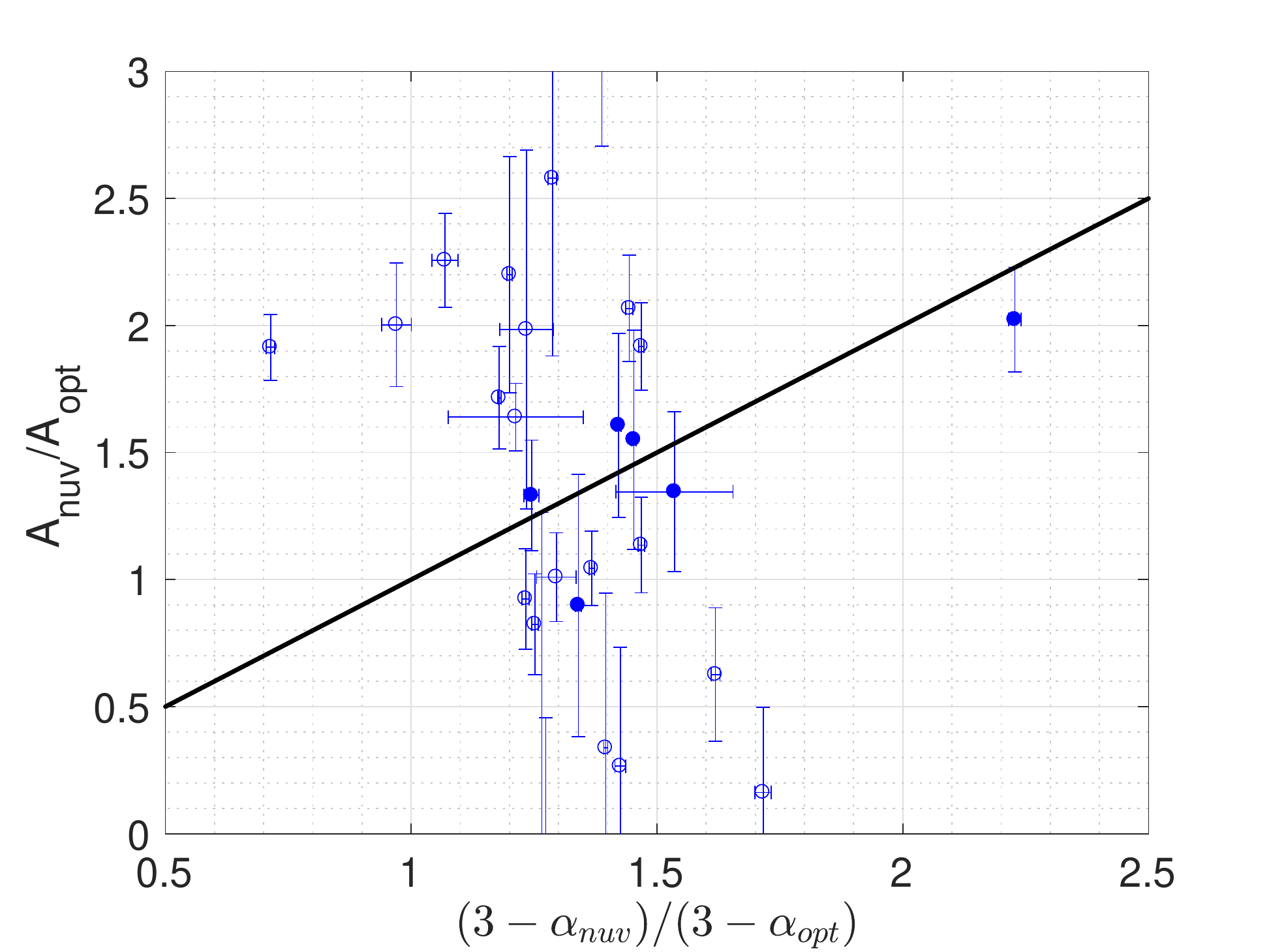}
\caption{The ratio of variability amplitudes in the nUV and optical
  bands, measured directly from the light curves {\it versus} the
  amplitudes expected from relativistic Doppler boost, calculated from
  the measured spectral indices in the two bands. The equality line
  corresponds to the Doppler-boost prediction from
  eq.~(\ref{eq:DooplerBoost}). The filled symbols indicate sources
  that are consistent with this Doppler signature within their
  1-$\sigma$ error.}
\label{Fig:DopplerBoostnUV}
\end{figure}

We tested the null hypothesis that an apparent Doppler boost signature
described by eq.~(\ref{eq:DooplerBoost}) arises by chance, given the
chromatic variability of quasars. For this test, we checked whether
the relative variability amplitudes measured directly from the light
curves (left side of the equation) are equal (within their 1-$\sigma$
uncertainty) to the relative amplitudes predicted from the Doppler
effect, which depends on the respective spectral slopes (right side of
the equation).  In Fig.~\ref{Fig:DopplerBoostnUV}, we show the
measured ratio of the nUV to the optical varibility amplitude, {\it
  versus} the Doppler-boost prediction calculated from the measured
spectral slopes, for the 30 quasars for which all necessary data were
available. We also show the equality line for comparison, and we
indicate the sources for which the variability is consistent with
eq.~(\ref{eq:DooplerBoost}) by filled symbols.
In total, in the nUV band, we found that the variability is consistent
with the Doppler prediction in 6 out of 30 quasars. This means that
the Doppler boost signature can be observed by chance in
$P(nUV)\equiv20_{-6}^{+8}$\%\footnote{We calculated the 1-$\sigma$
  confidence intervals with the Wilson (score) method \citep{wilson1927probable}.}
of the quasars in the sample.

\begin{figure}
\centering
\includegraphics[height=6cm,width=8cm]{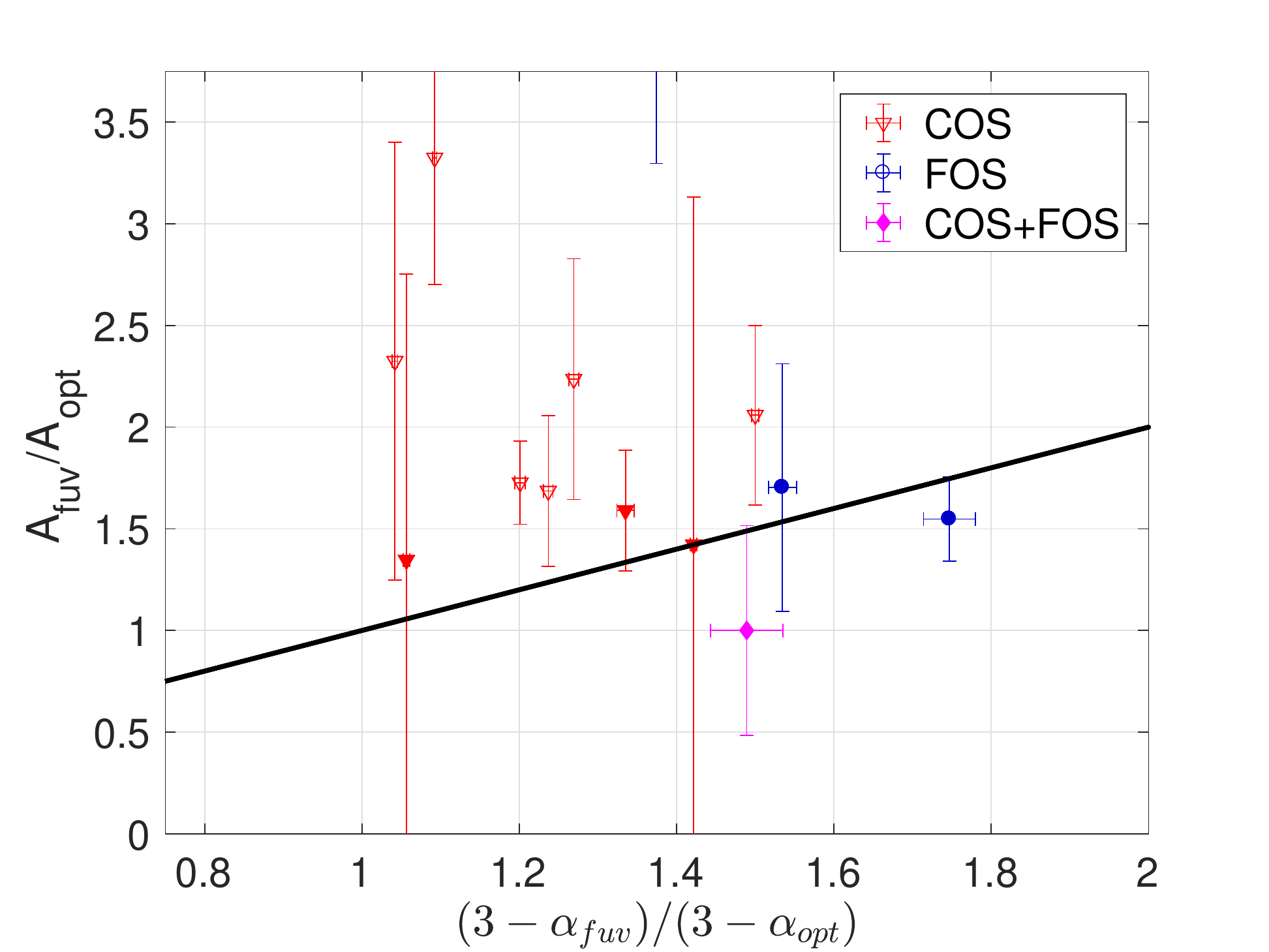}
\caption{Null hypothesis test for the Doppler boost signature, as in
  Fig.~\ref{Fig:DopplerBoostnUV}, but in the fUV band. Red triangles
  correspond to data from the COS catalog, blue circles to data from
  FOS and the purple diamond to the one AGN with spectra from both FOS
  and COS. As in Fig.~\ref{Fig:DopplerBoostnUV}, filled symbols
  illustrate the sources which satisfy eq.~(\ref{eq:DooplerBoost}).}
\label{Fig:DopplerBoostfUV}
\end{figure}

We next repeated the above test for quasars in the fUV band. In
Fig.~\ref{Fig:DopplerBoostfUV}, we show the amplitude ratio of fUV and
optical variability {\it versus} the value expected from a
Doppler boost in this band. Triangles correspond to data from the COS
catalog, circles to the data from FOS, and the diamond symbol marks
the one AGN which was included in both catalogs (the UV slope for this
source was calculated as the average slope from the two
spectra). Again, we indicate the sources for which variability is
consistent with eq.~(\ref{eq:DooplerBoost}) by filled symbols.
In this
case, we found that 6 out of 16 are consistent with
eq.~(\ref{eq:DooplerBoost}), i.e. an apparent Doppler signature can
arise by random chance $P(fUV)\equiv37.5_{-11.5}^{+12.5}$\% of the time.

We also note that the data are randomly scattered and do not show any
significant correlation. 
More importantly for our purposes, the overall trends in the data,
using either the fUV or the nUV bands, do not follow
eq.~(\ref{eq:DooplerBoost}), delineated with the equality line. 
In fact, the Pearson correlation coefficient for the nUV (fUV) sample is -0.23 (-0.69).
This is clearly illustrated in
Figures~\ref{Fig:DopplerBoostnUV}~and~\ref{Fig:DopplerBoostfUV},
and suggests that the cases which satisfy the Doppler prediction from
eq. (\ref{eq:DooplerBoost}) should be due to chance coincidence owing
to the limited and noisy data.
In other words, the relatively high
probability of the Doppler-like signature arising by chance
simply reflects limitations of the best available control
sample, and not necessarily the intrinsic properties of
chromatic AGN variability.  The clear conclusion is that
higher quality temporally coincident optical and UV
light curves and spectra, for a larger number of aperiodic
AGN, must be collected in the future.

\subsection{Doppler Boost in Periodic Sample}
\label{subsectiopn:Results_Periodic_Sample}
For a sub-sample of 68 periodic sources (55 from G15 and 13 from C16),
which had available UV light curves and optical spectra, we calculated
the UV spectral indices that would be required in order for their
variability to be consistent with relativistic Doppler boost. For all
the above sources, we calculated the spectral index in the nUV band,
while for 27 of those (20 from G15 and 7 from C16), we were also able
to calculate the spectral slope in the fUV band. Subsequently, we
compared the estimated UV indices, including their 1-$\sigma$ error,
with the observed distribution of spectral slopes in the respective
band in the control sample.

\begin{figure}
\centering
\includegraphics[height=6cm,width=9cm]{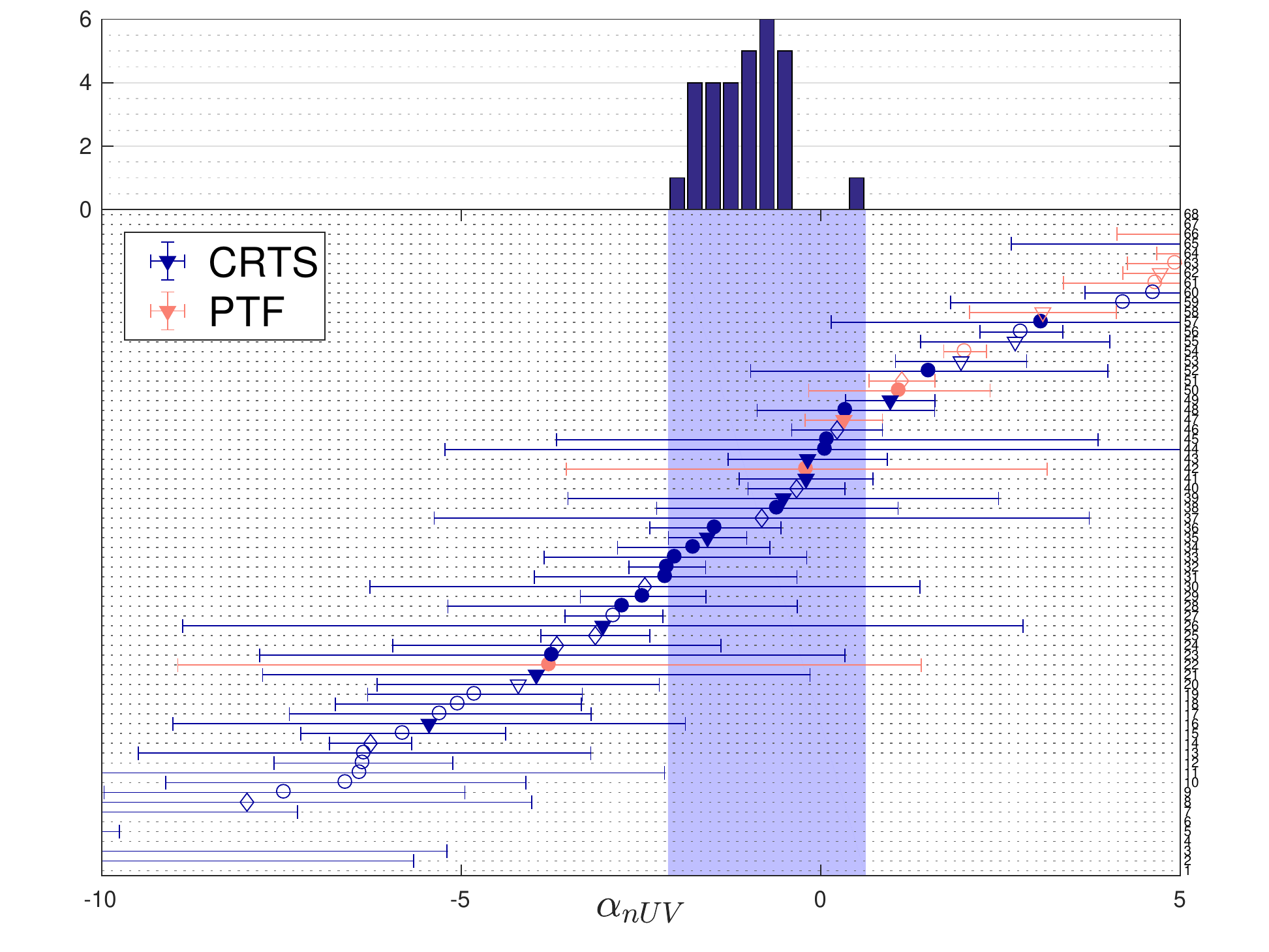}
\caption{{\em Top panel}: Distribution of nUV spectral indices
  measured from HST spectra for the sources in the control
  sample. {\em Bottom panel}: Inferred nUV spectral slopes for
  periodic quasars, assuming Doppler boost variability; filled (open)
  symbols for spectral indices consistent (inconsistent) with the
  observed distribution, circles when only the nUV spectral index was
  calculated, triangles when both nUV and fUV spectral slopes are
  calculated. We illustrate with open diamonds the sources the
  spectral indices of which are consistent with the observed slopes in
  one band but not in the other. Blue symbols present CRTS sources
  from G15 and red for PTF sources from C16. The indices on the right
  side of the plot refer to the entries in Table
  \ref{Table:BinaryProperties}. The shaded region delineates the range
  of the observed distribution shown in the top panel. }
\label{Fig:PeriodicSample_nUV}
\end{figure}

\begin{figure}
\centering
\includegraphics[height=6cm,width=9cm]{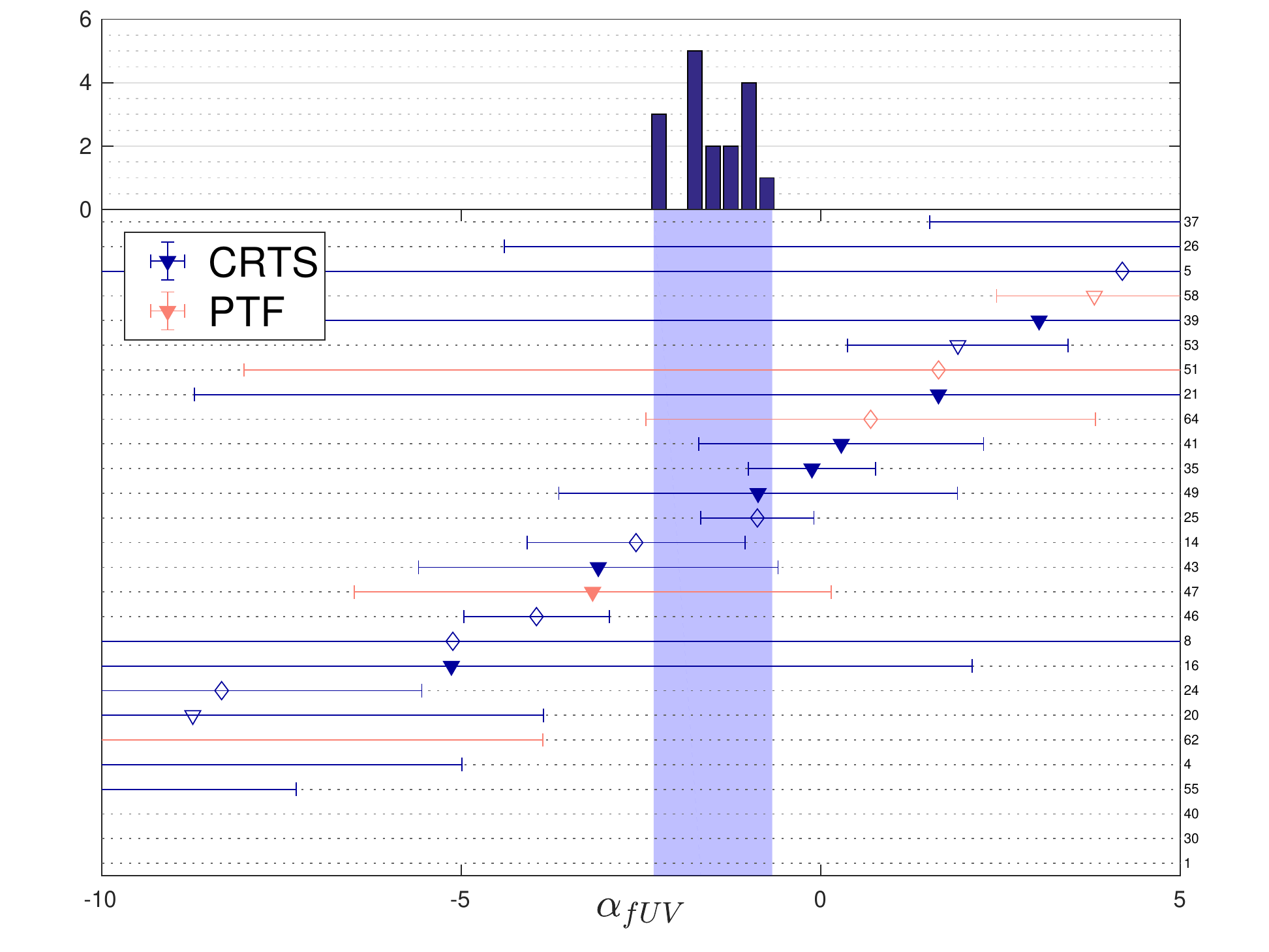}
\caption{Same as Fig.~\ref{Fig:PeriodicSample_nUV}, but in the fUV, instead of the nUV. 
All the sources in this sample have inferred spectral slopes in both UV bands.}
\label{Fig:PeriodicSample_fUV}
\end{figure}

In Figures~\ref{Fig:PeriodicSample_nUV}~and
~\ref{Fig:PeriodicSample_fUV}, we show the inferred nUV and fUV
slopes, respectively. We illustrate with circles the sources for which
we could calculate the spectral index only in the nUV band, whereas
with triangles we show the sources for which both spectral slopes
could be calculated. Filled symbols present sources with spectral
indices that are consistent within 1-$\sigma$ with the distribution of
spectral slopes observed in the control sample (filled circles: only
the nUV index was calculated and is consistent; filled triangles: both
indices were calculated and both were consistent with the observed
distributions) and open symbols present sources with indices that are
inconsistent with the observed distributions. In the case that both
slopes were determined, open triangles demonstrate sources that are
inconsistent in both bands, whereas open diamonds illustrate the
sources that are consistent in one band but not in the other. We
color-code the periodic sources from CRTS with blue symbols and the
sources from PTF with red. For clarity, we rank-ordered the sources
based on their inferred nUV and fUV slope, respectively. The numbers
on the right side of the figure refer to the entries in
Table~\ref{Table:BinaryProperties} below, in which we present the
details of the analysis for the periodic sample. On the top panel in
both Figures~\ref{Fig:PeriodicSample_nUV}~and
~\ref{Fig:PeriodicSample_fUV}, we show for comparison the distribution
of measured UV slopes from the control sample in the respective
band. The shaded regions in the bottom panels delineate the full
ranges of these distributions.

We found that 31 out of 68 sources (or $46_{-6}^{+6}\%$) are
consistent with the observed distribution if we consider only the nUV
slopes, and 15 out of 27 (or $55_{-9}^{+10}\%$) when considering only
the fUV spectral indices. If we require that both measured spectral
indices are consistent with their respective observed distributions,
then we find 9 out of 27 sources (or $33_{-8}^{+10}\%$) pass both
tests. These fractions are higher than expected based on the random
multi-wavelength variability found above.
This test, of course, can be used only to {\em rule~out} the Doppler
hypothesis for $\sim 2/3$rd of the sample.  Although the remaining
$\sim 1/3$rd of the sample remains consistent with the Doppler boost,
it is necessary to obtain their UV spectra, and directly measure their
spectral slopes, to check this.  Another caveat is that we are
comparing to the distribution observed in our small control sample;
the entire population of quasars likely has a wider distribution of
spectral slopes. Therefore, it is possible that we excluded some
sources that could be Doppler boosted.
In principle, this slope-distribution could be measured for many more
sources beyond the control sample we assembled, i.e objects with
available UV spectra but without a suitable UV light-curve.

Finally, we calculated whether the observed optical variability
amplitude is consistent with the maximum value allowed by a
relativistic Doppler boost (eq.~\ref{eq:FoM}).
We found that for 76 out of 94 sources from G15 and 18 out of 25 from
C16 the ratio is above unity and thus the Doppler boost model is
feasible. We show the estimated ratio in Table
\ref{Table:BinaryProperties}. 
We were able to exclude 5 additional sources, which, based on the spectral indices,
 could be consistent with the Doppler model  (filled
symbols in the Figures above), because they failed this criterion (i.e. the
ratio is below unity).
This clearly demonstrates that
tests of the Doppler model should not be limited to the relative
amplitudes from eq.~(\ref{eq:DooplerBoost}), but should require the
model to explain the overall optical and UV variability of the
sources to begin with.

\section{Discussion}
\label{sec:discussion}

\subsection{Expanding the sample size: UV spectral slope from photometric data?}
\label{subsection:UVSlopefromphotometry}

One of the main limitations of the current study is the small size of
the sample for which the test was feasible: all of the necessary
spectral and photometric information was available only for 42
sources. As mentioned above, this is primarily driven by the lack of
UV spectra.
On the other hand, for many of the sources without a UV spectrum,
photometric data exists in the two UV bands of GALEX. It is therefore
natural to ask if we could obtain reliable estimates of the UV
spectral slopes for these sources (e.g., by using spectral templates),
and thereby increase the sample size significantly.

Under the approximation that the continuum throughout the entire UV
band can be described by a single power-law, a single GALEX
observation with simultaneous photometric measurements in both the nUV
and the fUV bands determines the power-law slope. Such measurements are
available for a large number of quasars and AGN.  We assessed the
validity of this approach, using the sources in the control sample, by
calculating spectral indices from the photometry and comparing with
the measured slopes.

We first calculated the photometric flux in each band
(see \url{https://asd.gsfc.nasa.gov/archive/galex/FAQ/counts_background.html}),
\begin{align}
\label{eq:Galex_Flux}
F_{\rm nUV}&=2.06\times10^{-16}\times10^{\left( -\frac{m_{\rm nUV}-20.08}{2.5}\right)},
\\
F_{\rm fUV}&=1.40\times10^{-15}\times10^{\left(-\frac{m_{\rm fUV}-18.82}{2.5}\right)},
\end{align}
and subsequently the spectral slope in the UV band, assuming that the
fluxes are measured at the effective wavelengths $\lambda_{\rm
  fUV}^{\rm eff}=1542.26\AA$ and $\lambda_{\rm nUV}^{\rm
  eff}=2274.37\AA$, respectively.  If the UV light curve had multiple
simultaneous observations in both bands, we calculated an average of
the spectral slopes measured from each individual epoch.

\begin{figure}
\centering
\includegraphics[height=6cm,width=8cm]{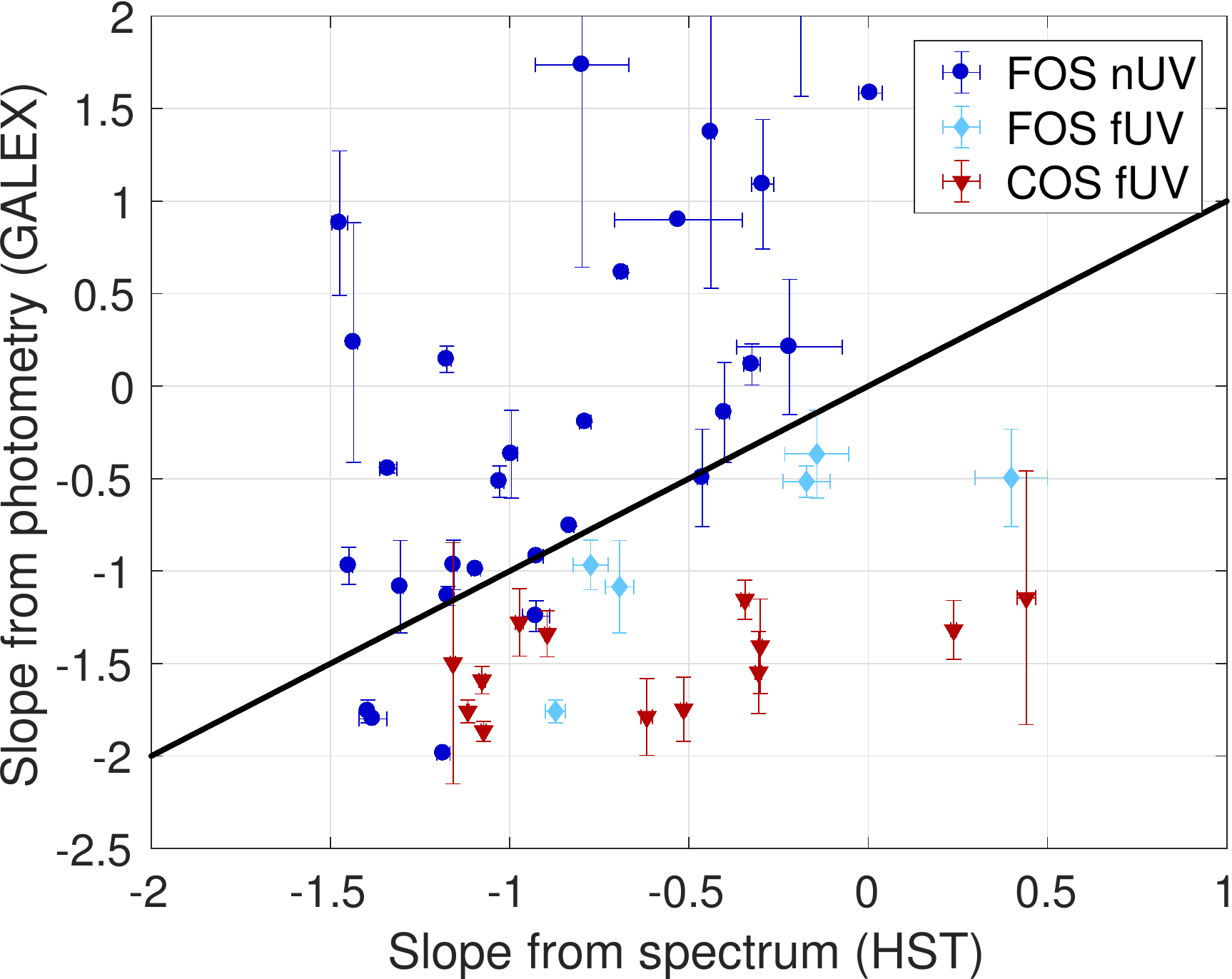}
\caption{Spectral slopes estimated from GALEX photometric data {\it
    versus} measured by fitting HST spectra with a power-law.  Dark
  blue circles: nUV spectra from FOS; light-blue diamonds: fUV spectra
  from FOS; red triangles: fUV spectra from COS.}
\label{Fig:PhotometryVsSpectrum}
\end{figure}

In Fig.~\ref{Fig:PhotometryVsSpectrum}, we present these photometric
estimates of the spectral indices {\it versus} the true slopes
measured directly from {\it HST} spectra. Dark blue circles and light
blue diamonds indicate data from the FOS catalog in the nUV and fUV
bands, respectively, and red triangles show fUV data from the COS
catalog. The equality line is also shown for comparison.
We see that the slopes inferred from photometry are not strongly
correlated with the slopes measured directly from the spectra. The
spectral slopes tend to be overestimated compared to the measured ones
in the nUV band and underestimated in fUV. Since the slope calculated
from photometry is roughly similar to the average of the spectral
slopes in the two bands, this trend is not surprising given that the
spectra tend to be steeper in the nUV band and relatively flatter in
fUV.\footnote{We note that the spectral slopes $\beta_{\lambda}$ are
  typically negative, as is also obvious from the figure.}  

While in principle, a more complicated spectral template (in place of
a single power-law) could account for this curvature, the large
scatter in Fig.~\ref{Fig:PhotometryVsSpectrum} clearly precludes
accurate photometric estimates.
The large scatter may be partly caused by the presence of broad
emission lines, which affect the calculation of the photometric fluxes
and thus the spectral indices. Depending on the redshift of the
source, a different set of lines are present in each of the broad-band
filters of GALEX and since our sample covers a wide redshift range, it
is complicated to account for this effect.

Another possibility would be to calculate the UV spectral indices from
spectral templates (e.g. \citealt{Shull+2012,Ivashchenko+2014}).
However, as shown in
Figs.~\ref{Fig:PeriodicSample_nUV}~and~\ref{Fig:PeriodicSample_fUV},
the observed distribution of measured spectral slopes is relatively
broad. Additionally, the variance of the spectral indices observed in
quasars with multiple spectra (mainly optical) is significant. Hence,
we expect that using a spectral template will not be sufficient for
this study, since it will not allow us to capture the spectral
variability of each quasar, which is crucial for our test.

For all the above reasons, we concluded that the expansion of the
sample using spectral slopes estimated from photometry is not
possible. However, we expect that the test will be feasible for a
larger sample of quasars in the near-future, once the co-added spectra
from STIS and the nUV spectra from COS become available in HSLA. A
complementary approach is to acquire UV spectroscopy for a sample of
quasars that already have well-sampled UV and optical light curves. In
fact, this may be the most time-efficient strategy to improve
the null hypothesis test presented here: acquiring well-sampled
UV/optical light curves for quasars that have UV and optical spectra
would require multi-year campaigns, whereas UV spectra (and
potentially optical spectra as well) for the already well-sampled
light curves could be obtained without a long wait.

\subsection{Null test in both UV bands}

In \S~\ref{section:Results_NullTest}, we showed that the probability
for the Doppler signature from eq.~(\ref{eq:DooplerBoost}) to occur by
random chance is $P({\rm nUV})=20_{-6}^{+8}$\% and $P({\rm
  fUV})=37.5_{-11.5}^{+12.5}$\% for the samples examined in the nUV and
fUV band, respectively. However, there are four sources in the control
sample, for which the null hypothesis test was feasible in both
bands. Of these, one is consistent with eq.~(\ref{eq:DooplerBoost}) in
both bands, one is not consistent in either band and the remaining two
are consistent with the Doppler signature only in the fUV band, but
not in the nUV band. From this sample, we can conclude that the
probability that the Doppler signature from
eq. (\ref{eq:DooplerBoost}) can arise by chance in both bands
simultaneously is $P({\rm nUV}\cap{\rm fUV})=25_{-15}^{+25}$\%. It is
also worth noting that the source which is consistent with the Doppler
signature in both bands is the only source which has two fUV spectra
from FOS and COS and the spectral index is calculated as the average
of the two. If we consider only the FOS spectrum (in the other three
cases, both the nUV and fUV slopes are calculated from the same
co-added spectra from FOS), the source is inconsistent with
eq.~(\ref{eq:DooplerBoost}) in the fUV.

As this last point illustrates, in this sub-sample, we are
particularly limited by small number statistics and sparse data; this
unfortunately precludes drawing conclusions from four sources about
the entire quasar population.
We note further that stochastic variations in the nUV and fUV spectral
slopes are likely to be correlated, and this correlation would need to
be quantified and taken into account, when performing a joint
null-test in the nUV and fUV bands.  This joint test would be
valuable, since the Doppler model, in principle, should account for
variability in both UV bands.  It would therefore be particularly
important to obtain simultaenous nUV and fUV data on a larger set of
objects.

\subsection{Variability amplitudes}

As mentioned above, another major limitation in the present study
arises from the very sparse UV light curves. Most of the GALEX light
curves have only 2-3 distinct epochs temporally coincident with the
optical baseline. This strongly affects the estimates of the relative
amplitudes, as is demonstrated by the large error bars in
Figs.~\ref{Fig:DopplerBoostnUV}~and~\ref{Fig:DopplerBoostfUV}. This
limitation is more severe in the fUV band, where the light curves
rarely sample more than two epochs, and the photometric uncertainties
are larger. In fact, the larger uncertainties in the relative
amplitudes, in combination with the smaller sample size, likely
explains why the Doppler signature can be observed by chance more
often in the fUV band (i.e. $\sim$20\% in nUV {\it vs.} $\sim$37\% in fUV).

Because of the small number of UV observations, our ability to
constrain the UV/optical amplitudes can also be limited by
``unfortunate" sampling. In order to illustrate this, consider a UV
light curve with only two epochs and the following extreme cases: (1)
the UV data points correspond to two epochs with very similar optical
magnitudes and (2) the UV data points trace the maximum and minimum
magnitude of the optical light curve. In the first case, it is almost
impossible to constrain the relative UV/optical amplitude, whereas in
the latter case, the amplitude is uniquely constrained. This is an
additional uncertainty incorporated into the large error bars in
Figs.~\ref{Fig:DopplerBoostnUV}~and~\ref{Fig:DopplerBoostfUV}.

Finally, for the available light curves, the optical and UV data were
not taken simultaneously. This further limits our ability to estimate
the relative variability amplitudes, because quasars show short-term
variability~(e.g.~\citealt{2009ApJ...698..895K}).\footnote{Short-term variability here 
refers to timescales of several days to a couple of months and not the variability between 1-10\,d
which \citealt{Caplar+2017} found to be overestimated due to the limited 
photometric precision of the time domain surveys.}
This is not included in our analysis, because our polynomial fits
effectively filter out the short-term variability of the optical light
curves. However, constructing light curves with simultaneous
measurements in optical and UV bands, which can be attained with an
instrument like the Ultraviolet/Optical Telescope (UVOT) on Swift, can
mitigate this issue and significantly improve our UV/optical
fits. Note that in the Doppler boost model, this short-term
variability can be explained only if the rest-frame luminosities of
the mini-discs fluctuate, e.g., due to variable accretion rate.

\subsection{Constant Spectral Slope}
\label{section:ConstantSpectralSlope}

Throughout our analysis, we assumed that the spectral slopes do not
evolve with time. An intrinsically time-variable spectral slope could,
in principle, be easily incorporated in the Doppler-boost
model. However, in practice, the assumption of a constant intrinsic
spectrum is necessary, because of the lack of multiple UV and optical
spectra for each source.\footnote{In the ideal case, we would test the
  Doppler signature with simultaneous UV and optical spectra.}  Since
the ultimate goal is to assess whether we can differentiate the
Doppler-boost variability from the intrinsic chromatic variability, it
is crucial to understand the importance of this limitation.

To investigate this issue, we consider the sample of quasars that have
multiple spectroscopic observations. There are seven sources with more
than one spectrum in the optical, and one source with two fUV
spectra. In Fig.~\ref{Fig:DopplerBoostnUV} and
\ref{Fig:DopplerBoostfUV}, these sources can be recognized from their
large horizontal error bars (the error in the spectral slope from a
single spectrum is typically small, as seen in the figures for the
other sources).  From these figures, we conclude that the error
introduced by a variable spectral slope is significantly smaller
compared to the uncertainty in constraining the relative amplitudes.

\subsection{Further improvements of the Doppler tests}

Since the chromatic variability of quasars does not appear correlated
with the Doppler prediction from eq.~(\ref{eq:DooplerBoost}), it is
reasonable to assume that with good quality data (a sufficient number
of data points, low photometric noise, simultaneous optical and UV
spectra) the variability caused by relativistic Doppler boost could be
distinguished much more easily from intrinsic variability.  In other
words, our main result, namely that the Doppler signature can arise by
chance as often as in $\sim$20\% ($\sim$37\%) of quasars in the nUV
(fUV) bands, likely reflects mainly the limited quality of the
currently available data.

A significant improvement could be achieved by increasing the size of
the control sample, especially if we can increase the number of
sources for which the null hypothesis can be tested simultaneously in
both the nUV and the fUV band. In this study, we assembled the control
sample starting from the sources that already had measured UV
spectra. However, as discussed in \S~\ref{section:Sample} the
requirement for UV light curves coincident with the optical baseline
was yet another major limitation on the size of the control sample.
Likewise, a further limitation results from the large error bars in
fitting the relative amplitudes of the optical and UV light curves,
mainly due to the small number of UV data points.  These limitations
could be addressed by first assembling a sample of quasars that have
well-sampled UV light-curves (in both nUV and fUV) temporally
coincident with existing optical light-curves, and then obtain UV
spectra for these sources, covering both bands.

Finally, in order to incorporate additional uncertainties arising from
the short-time variability of quasars, we could start from quasars
that have good quality UV spectra, and then construct well-sampled
optical and UV light curves with simultaneous observations. This can
be achieved with a telescope such as UVOT on the Swift
satellite. However, this would generally require a long wait; i.e. a
multi-year campaign to build up reasonably long baselines.

\section{Summary}
\label{sec:summary}

Relativistic Doppler boost is inevitable in compact SMBHBs, and
provides a unique multi-wavelength test for the binary nature of any
SMBHB candidate.  Here we examined this test in the context of binary
candidates identified as quasars with periodic optical variability.
Our main conclusions can be summarised as follows.

\begin{itemize}
\item We assembled a control sample of 42 quasars with aperiodic
  variability, and analysed it to test whether the Doppler boost
  signature is distinct from intrinsic chromatic brightness variability.
\item We found that in the best available control sample, variability
  consistent with the Doppler model can arise by chance for $\sim$20\%
  and $\sim$37\% of the quasars in the nUV and fUV band,
  respectively. The larger chance of coincidence is likely explained
  by the poorer quality of the fUV data.
\item For 68 SMBHB candidates identified as quasars with periodic
  variability, based on their optical/UV light curves and optical
  spectra, we calculated the UV spectral slopes that would be required
  for the periodicity to be caused by relativistic Doppler boost.
\item Of these, 26 sources could be explained by relativistic Doppler
  boost, since their inferred slopes are consistent with the observed
  range of spectral indices from the control sample.
\item We were additionally able to exclude 5 of the above sources,
  because even the most optimistic Doppler model cannot produce
  variability amplitudes as large as observed in the optical.
\end{itemize}

Overall, our paper suggests that quasars do not often mimic the
optical/UV colour-variations expected to arise from relativistic
Doppler boost.  We have also demonstrated the need for a larger and
better control sample.  With a sufficient number of UV data points,
low photometric noise, and simultaneous optical and UV spectra, such
an improved control sample could be constructed, and would allow an
improved test, conclusively ruling out that a Doppler colour signature
arises by chance from stochastic intrinsic variability.

\section*{Acknowledgement}
We thank Jules Halpern, Szabi Marka, Imre Bartos, Joe Lazio, Michele
Vallisneri and Matthew Graham for useful discussions and suggestions.
MC acknowledges support from the National Science Foundation (NSF)
NANOGrav Physics Frontier Center, award number 1430284.  Financial
support was provided by NASA through grants NNX15AB19G, NNX17AL82G,
and 16-SWIFT16-0015, and by NSF grant 1715661. Financial support to DJD was provided
from NASA through Einstein Postdoctoral Fellowship
award number PF6-170151. ZH also gratefully
acknowledges support from a Simons Fellowship in Theoretical Physics
and hospitality by NYU. 
\bibliography{ms}

\begin{thebibliography}{85}
\expandafter\ifx\csname natexlab\endcsname\relax\def\natexlab#1{#1}\fi

\bibitem[{{Amaro-Seoane} {et~al}\mbox{.}(2017){Amaro-Seoane}, {Audley},
  {Babak}, {Baker}, {Barausse}, {Bender}, {Berti}, {Binetruy}, {Born},
  {Bortoluzzi}, {Camp}, {Caprini}, {Cardoso}, {Colpi}, {Conklin}, {Cornish},
  {Cutler}, {Danzmann}, {Dolesi}, {Ferraioli}, {Ferroni}, {Fitzsimons}, {Gair},
  {Gesa Bote}, {Giardini}, {Gibert}, {Grimani}, {Halloin}, {Heinzel}, {Hertog},
  {Hewitson}, {Holley-Bockelmann}, {Hollington}, {Hueller}, {Inchauspe},
  {Jetzer}, {Karnesis}, {Killow}, {Klein}, {Klipstein}, {Korsakova}, {Larson},
  {Livas}, {Lloro}, {Man}, {Mance}, {Martino}, {Mateos}, {McKenzie},
  {McWilliams}, {Miller}, {Mueller}, {Nardini}, {Nelemans}, {Nofrarias},
  {Petiteau}, {Pivato}, {Plagnol}, {Porter}, {Reiche}, {Robertson},
  {Robertson}, {Rossi}, {Russano}, {Schutz}, {Sesana}, {Shoemaker}, {Slutsky},
  {Sopuerta}, {Sumner}, {Tamanini}, {Thorpe}, {Troebs}, {Vallisneri},
  {Vecchio}, {Vetrugno}, {Vitale}, {Volonteri}, {Wanner}, {Ward}, {Wass},
  {Weber}, {Ziemer}, \& {Zweifel}}]{eLISA2017}
{Amaro-Seoane} P. {et~al.}, 2017, ArXiv e-prints

\bibitem[{{Artymowicz} \& {Lubow}(1996)}]{AL96}
{Artymowicz} P., {Lubow} S.~H., 1996, \apjl, 467, L77+

\bibitem[{{Barnes}(2002)}]{Barnes2002}
{Barnes} J.~E., 2002, \mnras, 333, 481

\bibitem[{{Barnes} \& {Hernquist}(1992)}]{BarnesHernquist1992}
{Barnes} J.~E., {Hernquist} L., 1992, \araa, 30, 705

\bibitem[{{Begelman}, {Blandford} \& {Rees}(1980){Begelman}, {Blandford}, \&
  {Rees}}]{Begelman1980}
{Begelman} M.~C., {Blandford} R.~D., {Rees} M.~J., 1980, \nat, 287, 307

\bibitem[{{Bianchi} {et~al}\mbox{.}(2008){Bianchi}, {Chiaberge}, {Piconcelli},
  {Guainazzi}, \& {Matt}}]{2008MNRAS.386..105B}
{Bianchi} S., {Chiaberge} M., {Piconcelli} E., {Guainazzi} M., {Matt} G., 2008,
  \mnras, 386, 105

\bibitem[{{Bowen} {et~al}\mbox{.}(2017{\natexlab{a}}){Bowen}, {Campanelli},
  {Krolik}, {Mewes}, \& {Noble}}]{Bowen_2017b}
{Bowen} D.~B., {Campanelli} M., {Krolik} J.~H., {Mewes} V., {Noble} S.~C.,
  2017{\natexlab{a}}, \apj, 838, 42

\bibitem[{{Bowen} {et~al}\mbox{.}(2017{\natexlab{b}}){Bowen}, {Mewes},
  {Campanelli}, {Noble}, {Krolik}, \& {Zilhao}}]{Bowen_2017}
{Bowen} D.~B., {Mewes} V., {Campanelli} M., {Noble} S.~C., {Krolik} J.~H.,
  {Zilhao} M., 2017{\natexlab{b}}, ArXiv e-prints

\bibitem[{{Caplar}, {Lilly} \& {Trakhtenbrot}(2017){Caplar}, {Lilly}, \&
  {Trakhtenbrot}}]{Caplar+2017}
{Caplar} N., {Lilly} S.~J., {Trakhtenbrot} B., 2017, \apj, 834, 111

\bibitem[{{Charisi} {et~al}\mbox{.}(2016){Charisi}, {Bartos}, {Haiman},
  {Price-Whelan}, {Graham}, {Bellm}, {Laher}, \& {M{\'a}rka}}]{Charisi2016}
{Charisi} M., {Bartos} I., {Haiman} Z., {Price-Whelan} A.~M., {Graham} M.~J.,
  {Bellm} E.~C., {Laher} R.~R., {M{\'a}rka} S., 2016, \mnras, 463, 2145

\bibitem[{{Charisi} {et~al}\mbox{.}(2015){Charisi}, {Bartos}, {Haiman},
  {Price-Whelan}, \& {M{\'a}rka}}]{Charisi2015}
{Charisi} M., {Bartos} I., {Haiman} Z., {Price-Whelan} A.~M., {M{\'a}rka} S.,
  2015, \mnras, 454, L21

\bibitem[{{Colpi}(2014)}]{Colpi2014}
{Colpi} M., 2014, \ssr, 183, 189

\bibitem[{{Comerford} {et~al}\mbox{.}(2011){Comerford}, {Pooley}, {Gerke}, \&
  {Madejski}}]{2011ApJ...737L..19C}
{Comerford} J.~M., {Pooley} D., {Gerke} B.~F., {Madejski} G.~M., 2011, \apjl,
  737, L19

\bibitem[{{Cutri} {et~al}\mbox{.}(1985){Cutri}, {Wisniewski}, {Rieke}, \&
  {Lebofsky}}]{Cutri1985}
{Cutri} R.~M., {Wisniewski} W.~Z., {Rieke} G.~H., {Lebofsky} M.~J., 1985, \apj,
  296, 423

\bibitem[{{D'Orazio} \& {Haiman}(2017)}]{Dorazio2017}
{D'Orazio} D.~J., {Haiman} Z., 2017, ArXiv e-prints

\bibitem[{{D'Orazio} {et~al}\mbox{.}(2015{\natexlab{a}}){D'Orazio}, {Haiman},
  {Duffell}, {Farris}, \& {MacFadyen}}]{Dorazio2015}
{D'Orazio} D.~J., {Haiman} Z., {Duffell} P., {Farris} B.~D., {MacFadyen} A.~I.,
  2015{\natexlab{a}}, \mnras, 452, 2540

\bibitem[{{D'Orazio} {et~al}\mbox{.}(2015{\natexlab{b}}){D'Orazio}, {Haiman},
  {Duffell}, {MacFadyen}, \& {Farris}}]{Dorazio+2016}
{D'Orazio} D.~J., {Haiman} Z., {Duffell} P., {MacFadyen} A.~I., {Farris} B.~D.,
  2015{\natexlab{b}}, ArXiv e-prints

\bibitem[{{D'Orazio}, {Haiman} \& {MacFadyen}(2013){D'Orazio}, {Haiman}, \&
  {MacFadyen}}]{2013MNRAS.436.2997D}
{D'Orazio} D.~J., {Haiman} Z., {MacFadyen} A., 2013, \mnras, 436, 2997

\bibitem[{{D'Orazio}, {Haiman} \& {Schiminovich}(2015){D'Orazio}, {Haiman}, \&
  {Schiminovich}}]{Dorazio2015Nature}
{D'Orazio} D.~J., {Haiman} Z., {Schiminovich} D., 2015, \nat, 525, 351

\bibitem[{{D'Orazio} \& {Loeb}(2017)}]{Dorazio_2017_VLBI}
{D'Orazio} D.~J., {Loeb} A., 2017, ArXiv e-prints

\bibitem[{{Dotti}, {Sesana} \& {Decarli}(2012){Dotti}, {Sesana}, \&
  {Decarli}}]{Dotti+2012}
{Dotti} M., {Sesana} A., {Decarli} R., 2012, Advances in Astronomy, 2012,
  940568

\bibitem[{{Edelson}(1992)}]{Edelson1992}
{Edelson} R., 1992, \apj, 401, 516

\bibitem[{{Edelson} {et~al}\mbox{.}(1996){Edelson}, {Alexander}, {Crenshaw},
  {Kaspi}, {Malkan}, {Peterson}, {Warwick}, {Clavel}, {Filippenko}, {Horne},
  {Korista}, {Kriss}, {Krolik}, {Maoz}, {Nandra}, {O'Brien}, {Penton},
  {Yaqoob}, {Albrecht}, {Alloin}, {Ayres}, {Balonek}, {Barr}, {Barth},
  {Bertram}, {Bromage}, {Carini}, {Carone}, {Cheng}, {Chuvaev}, {Dietrich},
  {Dultzin-Hacyan}, {Gaskell}, {Glass}, {Goad}, {Hemar}, {Ho}, {Huchra},
  {Hutchings}, {Johnson}, {Kazanas}, {Kollatschny}, {Koratkar}, {Kovo}, {Laor},
  {MacAlpine}, {Magdziarz}, {Martin}, {Matheson}, {McCollum}, {Miller},
  {Morris}, {Oknyanskij}, {Penfold}, {Perez}, {Perola}, {Pike}, {Pogge},
  {Ptak}, {Qian}, {Recondo-Gonzalez}, {Reichert}, {Rodriguez-Espinoza},
  {Rodriguez-Pascual}, {Rokaki}, {Roland}, {Sadun}, {Salamanca}, {Santos-Lleo},
  {Shields}, {Shull}, {Smith}, {Smith}, {Snijders}, {Stirpe}, {Stoner}, {Sun},
  {Ulrich}, {van Groningen}, {Wagner}, {Wagner}, {Wanders}, {Welsh}, {Weymann},
  {Wilkes}, {Wu}, {Wurster}, {Xue}, {Zdziarski}, {Zheng}, \&
  {Zou}}]{Edelson1996}
{Edelson} R.~A. {et~al.}, 1996, \apj, 470, 364

\bibitem[{{Evans} \& {Koratkar}(2004)}]{Evans2004}
{Evans} I.~N., {Koratkar} A.~P., 2004, \apjs, 150, 73

\bibitem[{{Fabbiano} {et~al}\mbox{.}(2011){Fabbiano}, {Wang}, {Elvis}, \&
  {Risaliti}}]{2011Natur.477..431F}
{Fabbiano} G., {Wang} J., {Elvis} M., {Risaliti} G., 2011, \nat, 477, 431

\bibitem[{{Farris} {et~al}\mbox{.}(2014){Farris}, {Duffell}, {MacFadyen}, \&
  {Haiman}}]{Farris2014}
{Farris} B.~D., {Duffell} P., {MacFadyen} A.~I., {Haiman} Z., 2014, \apj, 783,
  134

\bibitem[{{Fu} {et~al}\mbox{.}(2015){Fu}, {Myers}, {Djorgovski}, {Yan},
  {Wrobel}, \& {Stockton}}]{2015ApJ...799...72F}
{Fu} H., {Myers} A.~D., {Djorgovski} S.~G., {Yan} L., {Wrobel} J.~M.,
  {Stockton} A., 2015, \apj, 799, 72

\bibitem[{{Giveon} {et~al}\mbox{.}(1999){Giveon}, {Maoz}, {Kaspi}, {Netzer}, \&
  {Smith}}]{Giveon1999}
{Giveon} U., {Maoz} D., {Kaspi} S., {Netzer} H., {Smith} P.~S., 1999, \mnras,
  306, 637

\bibitem[{{Gold} {et~al}\mbox{.}(2014){Gold}, {Paschalidis}, {Etienne},
  {Shapiro}, \& {Pfeiffer}}]{2014PhRvD..89f4060G}
{Gold} R., {Paschalidis} V., {Etienne} Z.~B., {Shapiro} S.~L., {Pfeiffer}
  H.~P., 2014, \prd, 89, 064060

\bibitem[{{Goulding} {et~al}\mbox{.}(2017){Goulding}, {Greene}, {Bezanson},
  {Greco}, {Johnson}, {Leauthaud}, {Matsuoka}, {Medezinski}, \&
  {Price-Whelan}}]{Goulding+2017}
{Goulding} A.~D. {et~al.}, 2017, ArXiv e-prints

\bibitem[{{Graham} {et~al}\mbox{.}(2014){Graham}, {Djorgovski}, {Drake},
  {Mahabal}, {Chang}, {Stern}, {Donalek}, \& {Glikman}}]{2014MNRAS.439..703G}
{Graham} M.~J., {Djorgovski} S.~G., {Drake} A.~J., {Mahabal} A.~A., {Chang} M.,
  {Stern} D., {Donalek} C., {Glikman} E., 2014, \mnras, 439, 703

\bibitem[{{Graham} {et~al}\mbox{.}(2015{\natexlab{a}}){Graham}, {Djorgovski},
  {Stern}, {Drake}, {Mahabal}, {Donalek}, {Glikman}, {Larson}, \&
  {Christensen}}]{Graham2015}
{Graham} M.~J. {et~al.}, 2015{\natexlab{a}}, \mnras, 453, 1562

\bibitem[{{Graham} {et~al}\mbox{.}(2015{\natexlab{b}}){Graham}, {Djorgovski},
  {Stern}, {Glikman}, {Drake}, {Mahabal}, {Donalek}, {Larson}, \&
  {Christensen}}]{Graham2015Nature}
{Graham} M.~J. {et~al.}, 2015{\natexlab{b}}, \nat, 518, 74

\bibitem[{{Green} {et~al}\mbox{.}(2010){Green}, {Myers}, {Barkhouse},
  {Mulchaey}, {Bennert}, {Cox}, \& {Aldcroft}}]{2010ApJ...710.1578G}
{Green} P.~J., {Myers} A.~D., {Barkhouse} W.~A., {Mulchaey} J.~S., {Bennert}
  V.~N., {Cox} T.~J., {Aldcroft} T.~L., 2010, \apj, 710, 1578

\bibitem[{{Haehnelt} \& {Kauffmann}(2002)}]{Haehnelt2002}
{Haehnelt} M.~G., {Kauffmann} G., 2002, \mnras, 336, L61

\bibitem[{{Haiman}, {Kocsis} \& {Menou}(2009){Haiman}, {Kocsis}, \&
  {Menou}}]{2009ApJ...700.1952H}
{Haiman} Z., {Kocsis} B., {Menou} K., 2009, \apj, 700, 1952

\bibitem[{{Hayasaki}, {Mineshige} \& {Sudou}(2007){Hayasaki}, {Mineshige}, \&
  {Sudou}}]{2007PASJ...59..427H}
{Hayasaki} K., {Mineshige} S., {Sudou} H., 2007, \pasj, 59, 427

\bibitem[{{Hung} {et~al}\mbox{.}(2016){Hung}, {Gezari}, {Jones}, {Kirshner},
  {Chornock}, {Berger}, {Rest}, {Huber}, {Narayan}, {Scolnic}, {Waters},
  {Wainscoat}, {Martin}, {Forster}, \& {Neill}}]{Hung2016}
{Hung} T. {et~al.}, 2016, \apj, 833, 226

\bibitem[{{Ivashchenko}, {Sergijenko} \& {Torbaniuk}(2014){Ivashchenko},
  {Sergijenko}, \& {Torbaniuk}}]{Ivashchenko+2014}
{Ivashchenko} G., {Sergijenko} O., {Torbaniuk} O., 2014, \mnras, 437, 3343

\bibitem[{{Jun} {et~al}\mbox{.}(2015){Jun}, {Stern}, {Graham}, {Djorgovski},
  {Mainzer}, {Cutri}, {Drake}, \& {Mahabal}}]{Jun2015}
{Jun} H.~D., {Stern} D., {Graham} M.~J., {Djorgovski} S.~G., {Mainzer} A.,
  {Cutri} R.~M., {Drake} A.~J., {Mahabal} A.~A., 2015, \apjl, 814, L12

\bibitem[{{Kauffmann} \& {Haehnelt}(2000)}]{KauffmanHaehnelt2000}
{Kauffmann} G., {Haehnelt} M., 2000, \mnras, 311, 576

\bibitem[{{Kelley}, {Blecha} \& {Hernquist}(2017){Kelley}, {Blecha}, \&
  {Hernquist}}]{Kelley+2017}
{Kelley} L.~Z., {Blecha} L., {Hernquist} L., 2017, \mnras, 464, 3131

\bibitem[{{Kelly}, {Bechtold} \& {Siemiginowska}(2009){Kelly}, {Bechtold}, \&
  {Siemiginowska}}]{2009ApJ...698..895K}
{Kelly} B.~C., {Bechtold} J., {Siemiginowska} A., 2009, \apj, 698, 895

\bibitem[{{Kinney} {et~al}\mbox{.}(1991){Kinney}, {Bohlin}, {Blades}, \&
  {York}}]{Kinney1991}
{Kinney} A.~L., {Bohlin} R.~C., {Blades} J.~C., {York} D.~G., 1991, \apjs, 75,
  645

\bibitem[{{Kocsis}, {Haiman} \& {Loeb}(2012{\natexlab{a}}){Kocsis}, {Haiman},
  \& {Loeb}}]{2012MNRAS.427.2680K}
{Kocsis} B., {Haiman} Z., {Loeb} A., 2012{\natexlab{a}}, \mnras, 427, 2680

\bibitem[{{Kocsis}, {Haiman} \& {Loeb}(2012{\natexlab{b}}){Kocsis}, {Haiman},
  \& {Loeb}}]{2012MNRAS.427.2660K}
{Kocsis} B., {Haiman} Z., {Loeb} A., 2012{\natexlab{b}}, \mnras, 427, 2660

\bibitem[{{Komossa} {et~al}\mbox{.}(2003){Komossa}, {Burwitz}, {Hasinger},
  {Predehl}, {Kaastra}, \& {Ikebe}}]{2003ApJ...582L..15K}
{Komossa} S., {Burwitz} V., {Hasinger} G., {Predehl} P., {Kaastra} J.~S.,
  {Ikebe} Y., 2003, \apjl, 582, L15

\bibitem[{{Komossa} \& {Zensus}(2016)}]{KomossaZensus2016}
{Komossa} S., {Zensus} J.~A., 2016, in IAU Symposium, Vol. 312, Star Clusters
  and Black Holes in Galaxies across Cosmic Time, {Meiron} Y., {Li} S., {Liu}
  F.-K., {Spurzem} R., eds., pp. 13--25

\bibitem[{{Kormendy} \& {Ho}(2013)}]{Kormendy2013}
{Kormendy} J., {Ho} L.~C., 2013, \araa, 51, 511

\bibitem[{{Koss} {et~al}\mbox{.}(2011){Koss}, {Mushotzky}, {Treister},
  {Veilleux}, {Vasudevan}, {Miller}, {Sanders}, {Schawinski}, \&
  {Trippe}}]{2011ApJ...735L..42K}
{Koss} M. {et~al.}, 2011, \apjl, 735, L42

\bibitem[{{Koz{\l}owski} {et~al}\mbox{.}(2010){Koz{\l}owski}, {Kochanek},
  {Udalski}, {Wyrzykowski}, {Soszy{\'n}ski}, {Szyma{\'n}ski}, {Kubiak},
  {Pietrzy{\'n}ski}, {Szewczyk}, {Ulaczyk}, {Poleski}, \& {OGLE
  Collaboration}}]{2010ApJ...708..927K}
{Koz{\l}owski} S. {et~al.}, 2010, \apj, 708, 927

\bibitem[{{Kun} {et~al}\mbox{.}(2015){Kun}, {Frey}, {Gab{\'a}nyi}, {Britzen},
  {Cseh}, \& {Gergely}}]{Kun2015}
{Kun} E., {Frey} S., {Gab{\'a}nyi} K.~{\'E}., {Britzen} S., {Cseh} D.,
  {Gergely} L.~{\'A}., 2015, \mnras, 454, 1290

\bibitem[{{Kun} {et~al}\mbox{.}(2014){Kun}, {Gab{\'a}nyi}, {Karouzos},
  {Britzen}, \& {Gergely}}]{Kun2014}
{Kun} E., {Gab{\'a}nyi} K.~{\'E}., {Karouzos} M., {Britzen} S., {Gergely}
  L.~{\'A}., 2014, \mnras, 445, 1370

\bibitem[{{Liu} {et~al}\mbox{.}(2016){Liu}, {Gezari}, {Burgett}, {Chambers},
  {Draper}, {Hodapp}, {Huber}, {Kudritzki}, {Magnier}, {Metcalfe}, {Tonry},
  {Wainscoat}, \& {Waters}}]{Liu2016}
{Liu} T. {et~al.}, 2016, \apj, 833, 6

\bibitem[{{Liu} {et~al}\mbox{.}(2015){Liu}, {Gezari}, {Heinis}, {Magnier},
  {Burgett}, {Chambers}, {Flewelling}, {Huber}, {Hodapp}, {Kaiser},
  {Kudritzki}, {Tonry}, {Wainscoat}, \& {Waters}}]{Liu2015}
{Liu} T. {et~al.}, 2015, \apjl, 803, L16

\bibitem[{{Lu} {et~al}\mbox{.}(2016){Lu}, {Li}, {Bi}, \& {Wang}}]{Lu+2016}
{Lu} K.-X., {Li} Y.-R., {Bi} S.-L., {Wang} J.-M., 2016, \mnras, 459, L124

\bibitem[{{MacFadyen} \& {Milosavljevi{\'c}}(2008)}]{2008ApJ...672...83M}
{MacFadyen} A.~I., {Milosavljevi{\'c}} M., 2008, \apj, 672, 83

\bibitem[{{MacLeod} {et~al}\mbox{.}(2010){MacLeod}, {Ivezi{\'c}}, {Kochanek},
  {Koz{\l}owski}, {Kelly}, {Bullock}, {Kimball}, {Sesar}, {Westman}, {Brooks},
  {Gibson}, {Becker}, \& {de Vries}}]{2010ApJ...721.1014M}
{MacLeod} C.~L. {et~al.}, 2010, \apj, 721, 1014

\bibitem[{{Madau} {et~al}\mbox{.}(1996){Madau}, {Ferguson}, {Dickinson},
  {Giavalisco}, {Steidel}, \& {Fruchter}}]{Madau1996}
{Madau} P., {Ferguson} H.~C., {Dickinson} M.~E., {Giavalisco} M., {Steidel}
  C.~C., {Fruchter} A., 1996, \mnras, 283, 1388

\bibitem[{{Manchester} \& {IPTA}(2013)}]{iPTA}
{Manchester} R.~N., {IPTA}, 2013, Classical and Quantum Gravity, 30, 224010

\bibitem[{{Mohan} {et~al}\mbox{.}(2016){Mohan}, {An}, {Frey}, {Mangalam},
  {Gab{\'a}nyi}, \& {Kun}}]{Mohan2016}
{Mohan} P., {An} T., {Frey} S., {Mangalam} A., {Gab{\'a}nyi} K.~{\'E}., {Kun}
  E., 2016, \mnras, 463, 1812

\bibitem[{{Mushotzky} {et~al}\mbox{.}(2011){Mushotzky}, {Edelson},
  {Baumgartner}, \& {Gandhi}}]{2011ApJ...743L..12M}
{Mushotzky} R.~F., {Edelson} R., {Baumgartner} W., {Gandhi} P., 2011, \apjl,
  743, L12

\bibitem[{{Noble} {et~al}\mbox{.}(2012){Noble}, {Mundim}, {Nakano}, {Krolik},
  {Campanelli}, {Zlochower}, \& {Yunes}}]{2012ApJ...755...51N}
{Noble} S.~C., {Mundim} B.~C., {Nakano} H., {Krolik} J.~H., {Campanelli} M.,
  {Zlochower} Y., {Yunes} N., 2012, \apj, 755, 51

\bibitem[{{Paltani} \& {Courvoisier}(1994)}]{Paltani1994}
{Paltani} S., {Courvoisier} T.~J.-L., 1994, \aap, 291, 74

\bibitem[{{Peeples} {et~al}\mbox{.}(2016){Peeples}, {Tumlinson}, {Fox},
  {Aloisi}, {Ayres}, {Danforth}, {Fleming}, {Jenkins}, {Jedrzejewski},
  {Keeney}, \& {Oliveira}}]{HLSA}
{Peeples} M.~S. {et~al.}, 2016, in American Astronomical Society Meeting
  Abstracts, Vol. 227, American Astronomical Society Meeting Abstracts, p.
  444.01

\bibitem[{{Rafikov}(2013)}]{2013ApJ...774..144R}
{Rafikov} R.~R., 2013, \apj, 774, 144

\bibitem[{{Rafikov}(2016)}]{2016arXiv160205206R}
{Rafikov} R.~R., 2016, ArXiv e-prints

\bibitem[{{Rodriguez} {et~al}\mbox{.}(2006){Rodriguez}, {Taylor}, {Zavala},
  {Peck}, {Pollack}, \& {Romani}}]{2006ApJ...646...49R}
{Rodriguez} C., {Taylor} G.~B., {Zavala} R.~T., {Peck} A.~B., {Pollack} L.~K.,
  {Romani} R.~W., 2006, \apj, 646, 49

\bibitem[{{Roedig} {et~al}\mbox{.}(2012){Roedig}, {Sesana}, {Dotti}, {Cuadra},
  {Amaro-Seoane}, \& {Haardt}}]{2012A&A...545A.127R}
{Roedig} C., {Sesana} A., {Dotti} M., {Cuadra} J., {Amaro-Seoane} P., {Haardt}
  F., 2012, \aap, 545, A127

\bibitem[{{Ruan} {et~al}\mbox{.}(2014){Ruan}, {Anderson}, {Dexter}, \&
  {Agol}}]{Ruan2014}
{Ruan} J.~J., {Anderson} S.~F., {Dexter} J., {Agol} E., 2014, \apj, 783, 105

\bibitem[{{Ryan} \& {MacFadyen}(2017)}]{Ryan2017}
{Ryan} G., {MacFadyen} A., 2017, \apj, 835, 199

\bibitem[{{Sakata} {et~al}\mbox{.}(2011){Sakata}, {Morokuma}, {Minezaki},
  {Yoshii}, {Kobayashi}, {Koshida}, \& {Sameshima}}]{Sakata2011}
{Sakata} Y., {Morokuma} T., {Minezaki} T., {Yoshii} Y., {Kobayashi} Y.,
  {Koshida} S., {Sameshima} H., 2011, \apj, 731, 50

\bibitem[{{Sesana} {et~al}\mbox{.}(2017){Sesana}, {Haiman}, {Kocsis}, \&
  {Kelley}}]{Sesana2017}
{Sesana} A., {Haiman} Z., {Kocsis} B., {Kelley} L.~Z., 2017, ArXiv e-prints

\bibitem[{{Shull}, {Stevans} \& {Danforth}(2012){Shull}, {Stevans}, \&
  {Danforth}}]{Shull+2012}
{Shull} J.~M., {Stevans} M., {Danforth} C.~W., 2012, \apj, 752, 162

\bibitem[{{Simm} {et~al}\mbox{.}(2016){Simm}, {Salvato}, {Saglia}, {Ponti},
  {Lanzuisi}, {Trakhtenbrot}, {Nandra}, \& {Bender}}]{2016A&A...585A.129S}
{Simm} T., {Salvato} M., {Saglia} R., {Ponti} G., {Lanzuisi} G., {Trakhtenbrot}
  B., {Nandra} K., {Bender} R., 2016, \aap, 585, A129

\bibitem[{{Tang}, {Haiman} \& {MacFadyen}(2018){Tang}, {Haiman}, \&
  {MacFadyen}}]{Tang+2018}
{Tang} Y., {Haiman} Z., {MacFadyen} A., 2018, ArXiv e-prints

\bibitem[{{Tang}, {MacFadyen} \& {Haiman}(2017){Tang}, {MacFadyen}, \&
  {Haiman}}]{Tang+2017}
{Tang} Y., {MacFadyen} A., {Haiman} Z., 2017, \mnras, 469, 4258

\bibitem[{{Vanden Berk} {et~al}\mbox{.}(2004){Vanden Berk}, {Wilhite}, {Kron},
  {Anderson}, {Brunner}, {Hall}, {Ivezi{\'c}}, {Richards}, {Schneider}, {York},
  {Brinkmann}, {Lamb}, {Nichol}, \& {Schlegel}}]{VandenBerk2004}
{Vanden Berk} D.~E. {et~al.}, 2004, \apj, 601, 692

\bibitem[{{Vaughan} {et~al}\mbox{.}(2016){Vaughan}, {Uttley}, {Markowitz},
  {Huppenkothen}, {Middleton}, {Alston}, {Scargle}, \& {Farr}}]{Vaughan2016}
{Vaughan} S., {Uttley} P., {Markowitz} A.~G., {Huppenkothen} D., {Middleton}
  M.~J., {Alston} W.~N., {Scargle} J.~D., {Farr} W.~M., 2016, \mnras, 461, 3145

\bibitem[{{Welsh}, {Wheatley} \& {Neil}(2011){Welsh}, {Wheatley}, \&
  {Neil}}]{Welsh2011}
{Welsh} B.~Y., {Wheatley} J.~M., {Neil} J.~D., 2011, \aap, 527, A15

\bibitem[{{Wilhite} {et~al}\mbox{.}(2005){Wilhite}, {Vanden Berk}, {Kron},
  {Schneider}, {Pereyra}, {Brunner}, {Richards}, \& {Brinkmann}}]{Wilhite2005}
{Wilhite} B.~C., {Vanden Berk} D.~E., {Kron} R.~G., {Schneider} D.~P.,
  {Pereyra} N., {Brunner} R.~J., {Richards} G.~T., {Brinkmann} J.~V., 2005,
  \apj, 633, 638

\bibitem[{Wilson(1927)}]{wilson1927probable}
Wilson E.~B., 1927, Journal of the American Statistical Association, 22, 209

\bibitem[{{Zheng} {et~al}\mbox{.}(2016){Zheng}, {Butler}, {Shen}, {Jiang},
  {Wang}, {Chen}, \& {Cuadra}}]{Zheng2016}
{Zheng} Z.-Y., {Butler} N.~R., {Shen} Y., {Jiang} L., {Wang} J.-X., {Chen} X.,
  {Cuadra} J., 2016, \apj, 827, 56

\bibitem[{{Zhu} {et~al}\mbox{.}(2016){Zhu}, {Wang}, {Cai}, \& {Sun}}]{Zhu2016}
{Zhu} F.-F., {Wang} J.-X., {Cai} Z.-Y., {Sun} Y.-H., 2016, \apj, 832, 75

\bibitem[{{Zu} {et~al}\mbox{.}(2013){Zu}, {Kochanek}, {Koz{\l}owski}, \&
  {Udalski}}]{2013ApJ...765..106Z}
{Zu} Y., {Kochanek} C.~S., {Koz{\l}owski} S., {Udalski} A., 2013, \apj, 765,
  106

\end{thebibliography}

\begin{table*}
\caption{Properties of periodic quasars analysed for relativistic Doppler boost.}
\label{Table:BinaryProperties}
\scriptsize
\begin{tabular}{|l| l| |c|c| c| c| c| c| c|}
\#& Name& $A_{\rm opt}$&$\alpha_{\rm opt}$&$A_{\rm nUV}/A_{\rm opt}$ &$\alpha_{\rm nUV}$&$A_{\rm fUV}/A_{\rm opt}$ &$\alpha_{\rm fUV}$ & $\frac{A_{\rm Dop}|_{max}}{2 A_{\rm opt}}$\\
\hline
1&SDSS J110554.78+322953.7&0.11$\pm$0.009&-1.56$\pm$0.09&5.91$\pm$0.71&-23.94$\pm$3.24&7.80$\pm$1.82&-32.56$\pm$8.29&1.62\\
2&SDSS J131706.19+271416.7&0.12$\pm$0.011&-0.27$\pm$0.12&7.61$\pm$4.96&-21.92$\pm$16.25&---&---&5.45\\
3&SDSS J072908.71+400836.6&0.06$\pm$0.005&-3.67$\pm$0.01&3.70$\pm$2.48&-21.72$\pm$16.52&---&---&0.65\\
4&3C 298.0&0.08$\pm$0.005&-1.15$\pm$0.01&5.18$\pm$0.72&-18.49$\pm$2.98&3.97$\pm$2.04&-13.47$\pm$8.48&6.44\\
5&SDSS J114438.34+262609.4&0.16$\pm$0.013&-1.16$\pm$0.01&4.12$\pm$1.05&-14.14$\pm$4.38&-0.29$\pm$5.10&4.19$\pm$21.20&3.18\\
6&SDSS J081133.43+065558.1&0.18$\pm$0.014&-1.37$\pm$0.01&3.26$\pm$0.28&-11.26$\pm$1.22&---&---&2.81\\
7&SDSS J153636.22+044127.0&0.09$\pm$0.008&-1.01$\pm$0.01&3.31$\pm$0.75&-10.30$\pm$3.02&---&---&3.20\\
8&BZQ J0842+4525&0.10$\pm$0.005&-1.46$\pm$0.01&2.46$\pm$0.89&-7.98$\pm$3.96&1.82$\pm$2.30&-5.12$\pm$10.27&5.34\\
9&SDSS J140704.43+273556.6&0.11$\pm$0.007&-0.58$\pm$0.01&2.92$\pm$0.70&-7.46$\pm$2.51&---&---&6.59\\
10&SDSS J130040.62+172758.4&0.30$\pm$0.022&-0.65$\pm$0.01&2.63$\pm$0.68&-6.61$\pm$2.50&---&---&0.86\\
11&SDSS J131909.08+090814.7&0.13$\pm$0.011&-0.90$\pm$0.01&2.41$\pm$1.09&-6.41$\pm$4.23&---&---&2.10\\
12&SDSS J125414.23+131348.1&0.16$\pm$0.012&-0.08$\pm$0.06&3.04$\pm$0.40&-6.37$\pm$1.24&---&---&1.42\\
13&SDSS J142301.96+101500.1&0.14$\pm$0.010&-1.29$\pm$0.05&2.18$\pm$0.73&-6.35$\pm$3.15&---&---&4.08\\
14&QNZ 3:54&0.19$\pm$0.013&-0.74$\pm$0.18&2.48$\pm$0.15&-6.26$\pm$0.57&1.49$\pm$0.41&-2.57$\pm$1.52&2.25\\
15&SDSS J104941.01+085548.4&0.16$\pm$0.009&-0.64$\pm$0.01&2.42$\pm$0.39&-5.81$\pm$1.43&---&---&2.86\\
16&\textbf{SDSS J082121.88+250817.5}&0.09$\pm$0.007&0.17$\pm$0.01&2.98$\pm$1.26&-5.45$\pm$3.56&2.87$\pm$2.56&-5.14$\pm$7.24&4.21\\
17&SDSS J081617.73+293639.6&0.14$\pm$0.012&-0.85$\pm$0.01&2.15$\pm$0.54&-5.29$\pm$2.10&---&---&4.51\\
18&SDSS J133654.44+171040.3&0.13$\pm$0.010&-1.14$\pm$0.01&1.94$\pm$0.41&-5.04$\pm$1.71&---&---&3.50\\
19&SDSS J102255.21+172155.7&0.13$\pm$0.014&-0.99$\pm$0.12&1.96$\pm$0.37&-4.81$\pm$1.49&---&---&1.79\\
20&SDSS J083349.55+232809.0&0.09$\pm$0.008&-1.21$\pm$0.01&1.71$\pm$0.47&-4.21$\pm$1.96&2.79$\pm$1.16&-8.74$\pm$4.88&6.11\\
21&\textbf{SDSS J104430.25+051857.2}&0.09$\pm$0.007&-1.05$\pm$0.01&1.72$\pm$0.94&-3.96$\pm$3.81&0.34$\pm$2.56&1.64$\pm$10.35&5.06\\
22*&\textbf{SDSS J170942.58+342316.2}&0.18$\pm$0.014&-1.56$\pm$0.01&1.48$\pm$1.13&-3.77$\pm$5.17&---&---&4.93\\
23&\textbf{HS 0926+3608}&0.08$\pm$0.006&-0.39$\pm$0.08&1.98$\pm$1.20&-3.73$\pm$4.07&---&---&9.11\\
24&RXS J10304+5516&0.08$\pm$0.008&-0.48$\pm$0.01&1.92$\pm$0.66&-3.67$\pm$2.28&3.26$\pm$0.80&-8.34$\pm$2.78&2.33\\
25&SDSS J121018.34+015405.9&0.12$\pm$0.009&-0.22$\pm$0.01&1.91$\pm$0.24&-3.14$\pm$0.76&1.21$\pm$0.24&-0.88$\pm$0.79&1.45\\
26&\textbf{SDSS J092911.35+203708.5}&0.21$\pm$0.014&-0.22$\pm$0.01&1.87$\pm$1.81&-3.03$\pm$5.84&-0.65$\pm$2.95&5.11$\pm$9.51&2.76\\
27&SDSS J224829.47+144418.0&0.32$\pm$0.019&0.39$\pm$0.01&2.25$\pm$0.26&-2.88$\pm$0.68&---&---&0.61\\
28&SDSS J104758.34+284555.8**&0.30$\pm$0.018&-0.32$\pm$0.01&1.73$\pm$0.73&-2.76$\pm$2.43&---&---&0.65\\
29&SDSS J144754.62+132610.0**&0.17$\pm$0.015&0.25$\pm$0.01&1.99$\pm$0.32&-2.47$\pm$0.87&---&---&0.85\\
30&SDSS J161013.67+311756.4&0.09$\pm$0.006&-0.99$\pm$0.01&1.36$\pm$0.96&-2.45$\pm$3.82&6.99$\pm$2.74&-24.93$\pm$10.94&1.38\\
31&\textbf{SDSS J094450.76+151236.9}&0.11$\pm$0.008&-0.18$\pm$0.08&1.62$\pm$0.57&-2.16$\pm$1.83&---&---&4.24\\
32&\textbf{SDSS J121457.39+132024.3}&0.20$\pm$0.012&-0.50$\pm$0.01&1.47$\pm$0.15&-2.14$\pm$0.53&---&---&2.09\\
33&SDSS J082926.01+180020.7**&0.22$\pm$0.015&-0.20$\pm$0.01&1.57$\pm$0.57&-2.02$\pm$1.89&---&---&0.84\\
34&\textbf{SDSS J154409.61+024040.0}&0.27$\pm$0.018&-1.25$\pm$0.01&1.12$\pm$0.25&-1.77$\pm$1.06&---&---&1.13\\
35&SDSS J140600.26+013252.2**&0.16$\pm$0.014&0.07$\pm$0.01&1.56$\pm$0.19&-1.58$\pm$0.54&1.07$\pm$0.30&-0.12$\pm$0.89&0.81\\
36&\textbf{SDSS J170616.24+370927.0}&0.13$\pm$0.012&-1.15$\pm$0.01&1.08$\pm$0.22&-1.47$\pm$0.91&---&---&2.99\\
37&SDSS J121018.66+185726.0&0.08$\pm$0.008&-0.95$\pm$0.01&0.97$\pm$1.15&-0.82$\pm$4.56&-2.47$\pm$2.85&12.77$\pm$11.25&6.52\\
38&\textbf{HS 1630+2355}&0.08$\pm$0.004&-0.78$\pm$0.01&0.95$\pm$0.44&-0.60$\pm$1.68&---&---&6.34\\
39&\textbf{SDSS J160730.33+144904.3}&0.12$\pm$0.010&0.06$\pm$0.01&1.20$\pm$1.02&-0.52$\pm$3.00&-0.01$\pm$3.48&3.04$\pm$10.21&4.24\\
40&SDSS J221016.97+122213.9&0.22$\pm$0.017&0.33$\pm$0.01&1.25$\pm$0.25&-0.34$\pm$0.68&7.85$\pm$1.93&-17.95$\pm$5.15&1.05\\
41&\textbf{SDSS J150450.16+012215.5}&0.14$\pm$0.009&-1.78$\pm$0.01&0.67$\pm$0.19&-0.22$\pm$0.93&0.57$\pm$0.41&0.28$\pm$1.98&3.58\\
42*&\textbf{SDSS J171617.49+341553.3}&0.14$\pm$0.012&-0.38$\pm$0.02&0.95$\pm$0.99&-0.20$\pm$3.34&---&---&6.07\\
43&\textbf{US 3204}&0.20$\pm$0.013&-0.80$\pm$0.01&0.84$\pm$0.29&-0.18$\pm$1.11&1.60$\pm$0.66&-3.10$\pm$2.50&1.61\\
44&\textbf{SDSS J133631.45+175613.8}&0.14$\pm$0.011&-0.14$\pm$0.02&0.94$\pm$1.69&0.06$\pm$5.29&---&---&1.66\\
45&\textbf{SDSS J135225.80+132853.2}&0.11$\pm$0.007&-0.13$\pm$0.02&0.93$\pm$1.20&0.09$\pm$3.77&---&---&1.61\\
46&UM 234&0.17$\pm$0.011&-0.34$\pm$0.01&0.83$\pm$0.19&0.23$\pm$0.63&2.08$\pm$0.30&-3.96$\pm$1.01&1.70\\
47*&\textbf{SDSS J231733.66+001128.3}&0.23$\pm$0.006&-0.38$\pm$0.01&0.79$\pm$0.15&0.32$\pm$0.54&1.82$\pm$0.97&-3.17$\pm$3.32&1.64\\
48&\textbf{SDSS J124157.90+130104.1}&0.21$\pm$0.015&-1.38$\pm$0.02&0.61$\pm$0.28&0.35$\pm$1.23&---&---&1.77\\
49&SDSS J082716.85+490534.0**&0.24$\pm$0.019&0.05$\pm$0.01&0.69$\pm$0.21&0.96$\pm$0.62&1.31$\pm$0.94&-0.87$\pm$2.77&0.94\\
50*&\textbf{SDSS J005158.83-002054.1}&0.21$\pm$0.009&0.08$\pm$0.01&0.65$\pm$0.43&1.09$\pm$1.26&---&---&1.53\\
51*&SDSS J212939.60+004845.5&0.21$\pm$0.014&0.04$\pm$0.01&0.63$\pm$0.16&1.13$\pm$0.46&0.46$\pm$3.26&1.64$\pm$9.67&2.93\\
52&\textbf{SDSS J103111.52+491926.5}&0.08$\pm$0.008&-1.21$\pm$0.01&0.36$\pm$0.59&1.51$\pm$2.48&---&---&4.83\\
53&SDSS J143820.60+055447.9&0.23$\pm$0.016&-0.01$\pm$0.01&0.35$\pm$0.30&1.95$\pm$0.91&0.36$\pm$0.51&1.91$\pm$1.53&0.50\\
54*&PDS 898&0.27$\pm$0.011&0.42$\pm$0.01&0.38$\pm$0.12&2.01$\pm$0.30&---&---&1.01\\
55&PGC 3096192&0.08$\pm$0.009&-0.39$\pm$0.01&0.09$\pm$0.39&2.70$\pm$1.32&5.86$\pm$2.83&-16.88$\pm$9.58&0.75\\
56&SDSS J084146.19+503601.1&0.19$\pm$0.013&0.61$\pm$0.01&0.09$\pm$0.24&2.79$\pm$0.58&---&---&0.28\\
57&\textbf{SDSS J164452.71+430752.2}&0.13$\pm$0.011&-1.20$\pm$0.07&-0.02$\pm$0.70&3.07$\pm$2.92&---&---&7.34\\
58*&UM 269&0.34$\pm$0.006&-0.52$\pm$0.01&-0.02$\pm$0.29&3.09$\pm$1.02&-0.23$\pm$0.39&3.80$\pm$1.36&0.69\\
59&SDSS J152157.02+181018.6&0.12$\pm$0.011&-0.64$\pm$0.01&-0.33$\pm$0.66&4.21$\pm$2.40&---&---&0.98\\
60&HS 0946+4845&0.12$\pm$0.007&0.01$\pm$0.01&-0.54$\pm$0.32&4.63$\pm$0.95&---&---&1.39\\
61*&SDSS J235958.72+003345.3&0.22$\pm$0.012&-1.43$\pm$0.01&-0.38$\pm$0.29&4.66$\pm$1.28&---&---&2.87\\
62*&SDSS J024442.77-004223.2&0.29$\pm$0.018&-0.81$\pm$0.04&-0.45$\pm$0.14&4.72$\pm$0.52&4.31$\pm$2.51&-13.42$\pm$9.55&0.87\\
63*&SDSS J141004.41+334945.5&0.21$\pm$0.006&0.26$\pm$0.01&-0.71$\pm$0.24&4.94$\pm$0.67&---&---&1.40\\
64*&SDSS J235928.99+170426.9&0.35$\pm$0.019&0.51$\pm$0.01&-1.07$\pm$0.40&5.66$\pm$0.99&0.93$\pm$1.26&0.69$\pm$3.13&0.99\\
65&SDSS J133516.17+183341.4&0.12$\pm$0.011&-0.86$\pm$0.01&-0.84$\pm$0.93&6.22$\pm$3.58&---&---&4.83\\
66*&SDSS J214036.77+005210.1&0.17$\pm$0.012&-0.34$\pm$0.24&-0.97$\pm$0.64&6.25$\pm$2.13&---&---&2.62\\
67&SDSS J123147.27+101705.3&0.24$\pm$0.016&-0.31$\pm$0.01&-2.18$\pm$1.21&10.23$\pm$4.00&---&---&1.38\\
68*&SDSS J171122.67+342658.9&0.33$\pm$0.020&0.06$\pm$0.01&-3.08$\pm$1.15&12.03$\pm$3.38&---&---&1.44\\

\end{tabular}
 \begin{flushleft}
We emphasize with bold the sources that are consistent with
relativistic Doppler boost.\\ * SMBHB candidates identified in PTF
(C16).\\ ** Sources that were consistent with the multi-wavelength
prediction of Doppler boost, but not with the maximum amplitude
requirement.
\end{flushleft}
\end{table*}

\end{document}